\documentclass[apj]{emulateapj}
\usepackage{apjfonts}
\usepackage{amsbsy}

\def\bb{$\bullet$}

\def\hmpc{$h^{-1}$Mpc}
\def\hkpc{$h^{-1}$kpc}

\def\hmsol{$h^{-1}$M$_\odot$}
\def\kms{km\,s$^{-1}$}

\def\om{\Omega_m}

\def\s8{\sigma_8}

\def\x2{$\chi^2$}
\def\hmsol{$h^{-1}\,$M$_\odot$}

\def\NNm1{\langle N(N-1) \rangle}

\def\m_star{M_\ast}

\def\om{\Omega_m}

\def\s8{\sigma_8}

\def\hmpc{$h^{-1}\,$Mpc}
\def\hkpc{$h^{-1}\,$kpc}

\def\x2{$\chi^2$}
\def\hmsol{$h^{-1}\,$M$_\odot$}

\def\kms{km\,s$^{-1}$}

\def\NNm1{\langle N(N-1) \rangle}

\def\p0{P_0(r)}

\newcommand{\cmjj}{\mbox{${\rm cm^{-2}}$}}
\newcommand{\hI}{\mbox{${\rm H\ I}$}}
\newcommand{\lya}{\mbox{${\rm Ly}\alpha$}}

\def\cg{\kappa_g}
\def\pb{s}

\def\m12{M_{12}}

\def\bb{\hat{b}_W}
\def\xf{\chi^2_f}
\def\xb{\chi^2_b}
\def\bhat{\hat{b}_W}
\def\wbar{\langle W_r\rangle}

\def\rshock{R_{\rm sh}}
\def\rg{R_g}%{R_{\rm gas}}
\def\rgas{R_g}%{R_{\rm gas}}
\def\m12{M_{12}}

\def\ashock{\gamma_{\rm sh}}%{\alpha_{\rm shock}}
\def\logmbar{\langle \log M\rangle}
\def\fcold{f_{\rm cold}}
\def\wr{W_r}
\def\aaa{\alpha}%{\alpha_{A1}}
\def\aab{\beta}%{\alpha_{A2}}
\def\fr{f(W_r)}
\def\rs0{\hat{R}_{\rm sh}^0}
\def\mg2{Mg\,II}

\begin{document}

\title{On The Halo Occupation of Dark Baryons}

\author{
Jeremy L. Tinker%\altaffilmark{1} 
\& Hsiao-Wen Chen%\altaffilmark{1}
}
\affil{Kavli Institute for Cosmological Physics \& Department of Astronomy and Astrophysics, University of Chicago}
%\altaffiltext{1}{Kavli Institute for Cosmological Physics, University of Chicago}
%\altaffiltext{2}{Department of Astronomy \& Astrophysics, University of Chicago}

%%%%%%%%%%%%%%%%%%%%%%%%%%%%%%%%%%%%%%%%%%%%%%%%%%%%%%%%%%%%%%%%%%%%%%%
% ABSTRACT ABSTRACT ABSTRACT ABSTRACT ABSTRACT ABSTRACT ABSTRACT 
%%%%%%%%%%%%%%%%%%%%%%%%%%%%%%%%%%%%%%%%%%%%%%%%%%%%%%%%%%%%%%%%%%%%%%%

\begin{abstract}

  We introduce a new technique that adopts the halo occupation
  framework for understanding the origin of QSO absorption-line
  systems.  Our initial study focuses specifically on Mg\,II
  absorbers.  We construct a model of the gaseous content in which the
  absorption equivalent width $\wr$ is determined by the the amount of
  cold gas, in the form of discrete clouds, along a sightline through
  a halo.  The two quantities that we specify per halo in the model
  are (1) the mean absorption strength per unit surface mass density
  $A_W(M)$, and (2) the mean covering factor $\cg(M)$ of the gaseous
  clouds.  These parameters determine the conditional probability
  distribution of $\wr$ as a function of halo mass, $P(\wr|M)$.  Two
  empirical measurements are applied to constrain the model: (i) the
  absorber frequency distribution function and (ii) the
  $\wr$-dependent clustering amplitude.  We find that the data demand
  a rapid transition in the gas content of halos at $\sim 10^{11.5}$
  \hmsol, below which halos contain predominantly cold gas and beyond
  which gas becomes predominantly hot.  In order to reproduce the
  observed overall strong clustering of the absorbers and the
  anti-correlation between $\wr$ and halo mass $M$, roughly 5\% of gas
  in halos up to $10^{14}$ \hmsol\ is required to be cold.  The gas
  covering factor is near unity over a wide range of halo mass,
  supporting that Mg\,II systems probe an unbiased sample of typical
  galaxies.  We discuss the implications of our study in the contexts
  of mass assembly of distant galaxies and the origin of QSO
  absorption line systems.

\end{abstract}

%%%%%%%%%%%%%%%%%%%%%%%%%%%%%%%%%%%%%%%%%%%%%%%%%%%%%%%%%%%%%%%%%%%%%%%
% NEW INTRODUCTION INTRODUCTION INTRODUCTION INTRODUCTION INTRODUCTION 
%%%%%%%%%%%%%%%%%%%%%%%%%%%%%%%%%%%%%%%%%%%%%%%%%%%%%%%%%%%%%%%%%%%%%%%

\section{Introduction}

The forest of \lya\ absorption line systems observed in the spectra of
background QSOs is a sensitive probe of intervening baryonic matter
that is otherwise invisible (see \citealt{rauch:98} for a
comprehensive review).  The gaseous clouds uncovered by these
absorption features span a wide range of ionization state and neutral
hydrogen column density over $N(\hI)= 10^{12}-10^{22}\,\cmjj$ and are
understood to originate in a range of environments in the dark matter
distribution, from moderate overdensities to fully collapsed
structures (e.g.\ \citealt{dave_etal:99}). Strong \lya\ absorbers with
$N(\hI)\gtrsim 2\times 10^{20}\,\cmjj$ are thought to be the
high-redshift analogue to present-day galaxies
(\citealt{wolfe_etal:05}). Studies of these high-density neutral gas
clouds show that they all contain low ions such as Si$^+$, Fe$^{+}$,
and Mg$^+$ (e.g.\ \citealt{lu_wolfe:94,
prochaska_etal:03}). Conversely, studies indicate that Mg\,II
absorbers of rest-frame absorption equivalent width $W_r(2796)\ge 0.3$
\AA\ arise in gaseous clouds of $N(\hI)= 10^{18}-10^{22}\,\cmjj$
(\citealt{churchill_etal:00, rao_etal:06}).  Over the redshift range
$z=0.35-2.5$, the Mg\,II $\lambda\lambda\,2796, 2803$ doublet
transitions are shifted to the optical spectral range $\lambda_{\rm
obs}=3800-10,000$ \AA\ and therefore provide a convenient optical
probe of dark baryons in extended gaseous halos of typical galaxies
along random lines of sight.

Extensive studies of the physical properties of Mg\,II absorbers have
been carried out by various authors.  Early studies that compared
abundance ratios between different ions associated with Mg\,II
absorbers showed that these absorbers arise in photo-ionized gas of
temperature $T\sim 10^4$ K (e.g.\ \citealt{bergeron_stasinska:86,
hamann:97}).  In addition, high-resolution spectra of strong absorbers
also show that these systems are multi-component, with $W_r$ roughly
proportional to the number of components in the system
(\citealt{petitjean_bergeron:90, churchill_vogt:01,
prochter_etal:06}).  Finally, comparisons of galaxies and Mg\,II
absorbers along common lines of sight also indicate that Mg\,II
absorbers at $\langle z\rangle=0.65$ are associated with luminous
galaxies (\citealt{bergeron:86, steidel_etal:94}) and that field
galaxies possess extended gaseous halos to projected distances of
$r\approx 50$ \hkpc\ (\citealt{lanzetta_bowen:90, steidel:93}).

Extended halos of cool gas around galaxies were predicted by
\cite{spitzer:56} and discussed in more detail in
\cite{bahcall_spitzer:69}. Theoretical models that explain the origin
of extended gaseous halos around galaxies include (i) the stripping of
gas from the accretion of small, gas-rich satellites into a larger
system (\citealt{wang:93}); (ii) cold gas gravitationally bound to
small-scale substructure within the dark matter halo (e.g.,
\citealt{sternberg_etal:02}); and (iii) a two-phase medium, cold
clouds condense out of a hotter halo through thermal instability
(\citealt{mo_jordi:96, maller_bullock:04, chelouche_etal:07}). These
models imply that $W_r$ reflects the potential well of the dark matter
halo---more massive systems sustain bigger gaseous halos with a higher
velocity dispersion.

However, recent studies have yielded conflicting results that
challenge this classical picture.  While some authors present
observations that suggest strong Mg\,II absorbers originating
primarily in galactic superwinds from low-mass galaxies
(\citealt{prochter_etal:06, bouche_etal:06, murphy_etal:07}), others
show that Mg\,II absorbing galaxies resemble typical field galaxies
(\citealt{zibetti_etal:05, kacprzak_etal:07, nestor_etal:07}) and the
covering fraction of Mg$^+$ ions is roughly 50\%
(\citealt{tripp_bowen:05}).  A particularly puzzling result is the
anti-correlation between $W_r(2796)$ (hereafter $\wr$) and clustering
amplitude by \cite{bouche_etal:06} (hereafter B06). These authors
measured the cross-correlation function of \mg2\ absorbers and
luminous red galaxies (LRGs) in data release three of the Sloan
Digital Sky Survey (SDSS DR3; \citealt{sdss_dr3}). From the relative
bias with respect to LRGs, B06 inferred the mean halo mass of
absorbers as a function of $W_r$, finding that strong absorbers of
$W_r=2-2.85$ \AA\ on average arise in less massive dark matter halos
of $\logmbar=11.11$, while absorbers of $W_r=0.3-1.15$ \AA\ on average
arise in dark matter halos of $\logmbar=12.49$. The authors consider
this anti-correlation in strong favor of a superwind origin for
$W_r>1$ \AA\ absorbers; the large equivalent width in low-mass systems
is attributed to the outflow velocity of the systems undergoing
starburst episodes.

The cross-correlation function of absorbers and galaxies on large
scales ($\gtrsim 1$\hmpc) provides a quantitative characterization of
the origin of these absorbers. The clustering of absorbers is a
consequence of the halos in which they are found. Low-mass halos form
in a wide range of environments while high-mass halos can only form in
the most dense regions of the dark matter distribution, thus high-mass
halos are highly clustered. The clustering of halos depends only on
simple gravitational physics, and the bias of dark matter halos as a
function of mass is a well-established relation in the standard CDM
cosmology. Thus it is straightforward to infer a mean halo mass from
large-scale clustering, as presented in B06.

However, when interpreting the clustering of gaseous clouds
based on the presence of a specific ion, it is necessary to consider
the state of the gas in the dark matter halos. For example, if the
observed Mg$^+$ ions originate in photo-ionized gas of temperature
$T\sim 10^4$ K, then the observed anti-correlation between $W_r$
and $\langle M\rangle$ may be produced when the halo gas in massive
dark matter halos becomes too hot for abundant Mg$^+$ to survive.  To
investigate how the content of warm/cold gas depends on halo mass, we
introduce a new technique that adopts the halo occupation framework
for studying the origin of Mg\,II absorbers.  This same technique can
also be applied for studying the nature of QSO absorption-line
systems.

Our halo occupation approach is similar to the Halo Occupation
Distribution (HOD) method that was developed to establish an empirical
mapping between galaxies and dark matter halos (e.g., \citealt{seljak:00,
  roman_etal:01, berlind_weinberg:02}; see \citealt{zheng_weinberg:07} and
references therein). This mapping is purely statistical in nature; it
interprets galaxy bias based on the probability that a halo of mass
$M$ contains $N$ galaxies of a given class, $P(N|M)$.  The
probability, $P(N|M)$, is constrained by measurements of the space
density of the class of objects and their two-point auto-correlation
function.  HOD analysis of galaxy clustering data can provide insight
into galaxy formation and evolution (\citealt{vdb_etal:03b, cooray:06,
  zheng_etal:07, white_etal:07, tinker_etal:07c}) and to constrain
cosmological parameters (\citealt{vdb_etal:03, vdb_etal:07,
  tinker_etal:05, tinker:07, zheng_weinberg:07}). In this paper, we
extend this framework to characterize the cold gas detected through
QSO absorption lines and focus specifically on Mg\,II absorbers.

To study the dark matter halo population of Mg\,II absorbers, we adopt
a conditional probability distribution of equivalent widths $P(W_r|M)$
to describe the halo occupation of cold gas.  Our technique is
fundamentally the same as the conditional luminosity function for
interpreting galaxy bias (\citealt{yang_etal:03, vdb_etal:07}).  In
our approach, we choose to parameterize $P(\wr|M)$---the distribution
of $\wr$ within a halo of given mass---according to expectations of a
uniform gaseous halo, in which the equivalent width is proportional to
the density-weighted path-length along a sight-line through the halo.
The parameters of $P(\wr|M)$ are left free and constrained by
empirical data.  The available observations are (1) the frequency
distribution function $d^2N/(d\wr\,dz)$, the number density of
absorbers per unit absorption equivalent width interval per unit
redshift path, and (2) the large-scale bias relative to LRGs.  The
halo occupation analysis allows us to go beyond simply estimating the
mean mass of dark matter halos that host absorbers and constrain the
distribution of these halos as a function of $\wr$.

We will consider two scenarios for $P(\wr|M)$. In the first scenario,
the fraction of cold gas relative to dark matter mass is a smoothly
varying function of halo mass. This scenerio is motivated by the
canonical model for halo gas in which accreting gas is always
shock-heated to the virial temperature when it crosses the virial
radius (e.g., \citealt{white_frenk:91}). The gas then cools from the
inside out, creating a cooling radius that varies with the properties
of the halo. We will demonstrate that this model cannot simultaneously
fit the frequency and bias data; models of this type that reproduce
the frequency distribution predict a positive correlation between
$\wr$ and $M$ because more massive halos are expected to contain more
absorbing gas.  Taking into account the fact that the Mg\,II
transitions probe primarily warm, photo-ionized gas, we include in the
second model a transitional mass scale, at which shock heated gas
progressively takes over higher mass halos and the absorption
efficiency is significantly reduced. In the transition region between
entirely `cold-mode' and `hot-mode' halos, heating occurs as an {\it
  inside-out} process.  Namely, a highly ionized core expands until it
envelopes the entire gas halo (e.g., \citealt{dekel_birnboim:06}).  We
allow the mass scale and width of this transition to be free
parameters.  In addition, we allow some fraction of cold gas in the
hot halos that is also constrained by empirical data.  We demonstrate
that the second scenario reproduces both the observed
$d^2N/(d\wr\,dz)$ and bias data well.  We obtain a best-fit
transitional mass scale of $\sim 10^{11.5}$\hmsol\ and a cold gas
fraction (as represented by the Mg$^+$ ions) of $< 10$\% in massive
halos.

Given the agreement between observations and our halo occupation
model, we argue that it is not necessary to invoke additional,
complicated star formation feedback for explaining the inverse
correlation between $W_r$ and $\langle M\rangle$.  We demonstrate that
the halo occupation approach allows us to gain physical insights for
(1) understanding the origin of intervening absorbers observed in the
spectra of background QSOs, and (2) constraining the distribution of
baryons in dark matter halos.  We discuss these results in the context
of gas accretion in recent theoretical developments (i.e.,
\citealt{birnboim_dekel:03, dekel_birnboim:06, keres_etal:05,
  birnboim_etal:07}).

This paper is organized as the following.  In \S\ 2, we outline the
theoretical framework for constructing the conditional probability
distribution of equivalent widths $P(\wr|M)$.  In \S\ 3, we derive the
observables that characterize the statistical properties of Mg\,II
absorbers, based on known dark matter halo statistics and
$P(\wr|M)$. In \S\ 4, we compare the model predictions with empirical
measurements for \mg2\ absorbers.  In \S\ 5, we apply the best-fit
model to predict additional statistical quantities on the
galaxy--absorber correlation that can be tested with future
observations.  Finally, we discuss in \S\ 6 the interpretations of our
analysis.  We adopt a flat $\Lambda$CDM cosmology with $\Omega_{\rm
M}=0.25$ and $\sigma_8 = 0.8$ and a dimensionless Hubble constant $h =
H_0/(100 \ {\rm km} \ {\rm s}^{-1}\ {\rm Mpc}^{-1})$ throughout the
paper.

%%%%%%%%%%%%%%%%%%%%%%%%%%%%%%%%%%%%%%%%%%%%%%%%%%%%%%%%%%%%%%%%%%%
% SECTION TWO 
%%%%%%%%%%%%%%%%%%%%%%%%%%%%%%%%%%%%%%%%%%%%%%%%%%%%%%%%%%%%%%%%%%%

\section{Theoretical Framework}

In this section, we describe the approach to populate dark matter
halos with cold baryons that are represented by the presence of Mg$^+$
ions.  An implicit assumption of our halo occupation approach is that
all absorbers originate in halos, rather than in metal enriched
intergalactic medium (IGM) outside of individual halos.  This
assumption is justified by the strong correlation between absorbers
and galaxies; the LRG--Mg\,II cross-correlation function exhibits a
strong signal at separations less than halo radii
(\citealt{bouche_etal:04}).  In addition, for each
absorber a galaxy is nearly always found at a separation significantly
less than the typical halo radius expected given the galaxy
luminosity.  Finally, the multi-component nature of \mg2\ systems
demonstrates that the absorbing medium is highly clumped, in contrast
to the smoother gas in the IGM.

For each dark matter halo, we will specify two quantities: (1)
the mean absorption strength per unit surface mass density $A_W(M)$,
and (2) the mean covering factor $\cg(M)$ of the gaseous clouds.
These parameters determine the conditional probability distribution of
$\wr$ as a function of halo mass, $P(\wr|M)$ which, when combined with
known statistics of dark matter halos, allows us to derive theoretical
predictions of various statistical properties of Mg\,II absorbers for
direct comparisons with observations.

%%%%%%%%%%%%%%%%%%%%%%%%%%%
\subsection{Constructing the Gaseous Halo}
%%%%%%%%%%%%%%%%%%%%%%%%%%%

A dark matter halo is defined as a collapsed and virialized object
with a mean interior density equal to 200 times the background
density. Thus the radius of a halo is
\begin{equation}
\label{e.R200}
R_h = R_{200} = \left[\frac{3M}{4\pi\, (200\,\bar{\rho}_m)}\right]^{1/3}.
\end{equation}
The choice of $200\,\bar{\rho}_m$ is motivated by estimates of the
virial radius from the spherical collapse model, which predict a value
of $\sim 180$. While these estimates are idealized, they also match
the halos identified in N-body simulations, from which the global
properties of halos are calibrated.  According to Equation
(\ref{e.R200}), a $10^{12}$ \hmsol\ halo has a co-moving radius of
$243$ \hkpc, while $R_{200}$ for a $10^{14}$ \hmsol\ halo is 1.1
\hmpc. With this definition, all halos of mass $M$ have the same
co-moving radius at any redshift. We adopt the halo mass profile of
\cite{nfw:97} and the concentration-mass relation of
\cite{bullock_etal:01} (with updated parameters from
\citealt{wechsler_etal:06}) to calculate the halo mass as a function
of radius.

We begin with the empirical result that the equivalent width $\wr$ of
an absorption system is proportional to the number of absorption
components, which we interpret as gas clouds or clumps encountered
along the sightline through the halo \footnote{We note that an
  absorption line of width $\wr$ implies a maximum velocity difference
  between components of $\sim 100$ \kms\ per Angstrom. The maximum circular
  velocity of a $10^{12}$ \hmsol\ NFW halo is $\sim 150$ \kms.}.
%\footnote{We note that an
%absorption line with a width of 2 \AA\ implies at maximum velocity
%difference of $\sim 200$ \kms\ between individual components, which is
%nearly twice the virial velocity of a $10^{11.5}$ \hmsol\ NFW halo,
%220 \kms. }.  
We adopt an isothermal profile with a core radius $a_h$
to approximate the mean gas density that follows
\begin{equation}
\label{e.si_sphere}
%\hat{\rho}_g(r) = \frac{M_h/4\pi}{R_g-a_h\tan^{-1}(R_g/a_h)}\left(r^2 + a_h^2\right)^{-1}
\rho_g(r) = f_g{\cal G}_0\left(r^2 + a_h^2\right)^{-1},
\end{equation}
where $f_g$ is the gas fraction,
\begin{equation}
\label{e.rho0}
{\cal G}_0 = \frac{M(<R_g)/4\pi}{R_g-a_h\tan^{-1}(R_g/a_h)},
\end{equation}
and $M(<\rg)$ is the mass of dark matter within the effective radius
$\rg$ of the gaseous halo.  Beyond $\rg$, no absorption is found.  We
note that an isothermal distribution of discrete gas clumps is
expected to have a mean absorption strength that declines with
radius. At a certain impact parameter the mean number of clouds per
line of sight will rapidly decline and become negligible (see, e.g.,
Figure 1 in \citealt{chelouche_etal:07}). We approximate this effect
by creating an effective gas radius, $\rg$, beyond which no gas clumps
are present.

Considering the total absorption equivalent width $\wr$ as a sum over
all clumps encountered along a sightline leads to
\begin{eqnarray}
\label{e.path_length}
\label{e.w_rho}
\wr(\pb|M) &=& W_0\times\left[\frac{2\sigma_{\rm cl}}{M_{\rm cl}}\int_0^{\sqrt{R_g^2-\pb^2}} {\rho}_g(\sqrt{\pb^2 + l^2})\,dl\right] \\
           &=& \left(\frac{W_0\,\sigma_{\rm cl}\,f_g}{M_{\rm cl}}\right) \frac{2{\cal G}_0}{\sqrt{\pb^2 + a_h^2}}\tan^{-1}\sqrt{\frac{R_g^2 - \pb^2}{\pb^2 + a_h^2}},
\end{eqnarray}
where $s$ is the impact parameter of the sightline with respect to the
center of the halo, $\sigma_{\rm cl}$ and $M_{\rm cl}$ are the cross
section and mean mass of individual gas clumps, and $W_0$ is the
absorption per clump.  It is clear that the parameters that control
the relation between impact parameter and $\wr$, such as the gas
fraction $f_g$, and the mean absorption strength $W_0$ and
cross-section $\sigma_{\rm cl}$ of individual clumps, are degenerate.
We express their product as a single free parameter
\begin{equation}
\label{e.a_w_defn}
A_W\equiv \frac{W_0\,\sigma_{\rm cl}\,f_g}{M_{\rm cl}},
\end{equation}
which represents the mean absorption equivalent width per unit surface
mass density of the cold gas. In the limit that all the components of
the absorption line are optically thin, $A_W$ would be a measure of
the mean particle density of Mg$^+$ ions. Some fraction of the lines
will be optically thick, however, so we are unable to constrain the
density from the values of $A_W$. The relation between the predicted
absorption equivalent width $W_r$ at impact parameter $s$ of a given
halo is described by
\begin{equation}
\label{e.w_rho2}
\wr(\pb|M) = A_W(M)\frac{2 {\cal G}_0}{\sqrt{\pb^2 + a_h^2}}\tan^{-1}\sqrt{\frac{R_g^2 - \pb^2}{\pb^2 + a_h^2}}.
\end{equation}
Equation (\ref{e.w_rho2}) shows that if the fraction of cold gas and
its properties do not change across all halo masses, then $A_W$ is
expected to be a constant and the total mean equivalent width is
expected to increase with halo mass due to the larger gas mass in more
massive halos.

We choose a value of $a_h=0.2\rg$, noting that our results are
insensitive to the exact choice for this parameter.  We set $\rg = 80$
comoving \hkpc\ (50 \hkpc\ physical at $z=0.6$) for halos of mass
$10^{12}$ \hmsol.  The gas radius scales as $\rg = 80
(M/10^{12})^{1/3}$ \hkpc, the same scaling as the halo radius in
equation (\ref{e.R200}). With this normalization, the gas radius is
roughly one third of the halo radius for all halo masses.  The choice
of $50$ \hkpc\ for our fiducial radius is motivated by observations of
the radial extent to which \mg2\ absorption of $\wr\ge 0.3$ \AA\ is
detected around $L_\ast$ galaxies (e.g., \citealt{steidel:93}). We
will demonstrate in \S\ 5 that the value of $\rgas$ makes distinct
predictions for the distribution of impact parameters, and that our
choice for $\rgas$ matches available observations well.

%%%%%%%%%%%%%%%%%%%%%%%%%%%%%%%%%%%%%%%%%%%%%%%%%%%%%%%%%%%%%%%%
\subsection{The Conditional Probability Distribution of $\wr$---A Smooth Gaseous Halo Model}
%%%%%%%%%%%%%%%%%%%%%%%%%%%%%%%%%%%%%%%%%%%%%%%%%%%%%%%%%%%%%%%%

To construct the conditional probability distribution of $\wr$,
$P(\wr|M)$, we consider the probability distribution of a random line
of sight intercepting a halo at impact parameter $\pb$ and the mean
absorption expected at the impact parameter $\pb$.  For a random line
of sight, the probability that a sightline passes through a halo of
mass $M$ with impact parameter $\pb$ is
\begin{equation}
\label{e.prho}
P(\pb|M) = \left\{ \begin{array}{ll}
		2\pb/ \rg^2 & {\rm if\ \ } \pb \le \rg \\
		0 & {\rm if\ \ } \pb > \rg . \\
		\end{array}
	\right.
\end{equation}
The distribution of $\wr$ is related to the distribution of
impact parameters according to
\begin{equation}
\label{e.pW_prhob}
P(\wr|M)d\wr = \cg(M)P(\pb|M)d\pb,
\end{equation}
where $\cg$ is the total probability of detecting an absorber in a
halo of mass $M$, which may be less than unity.  This integrated
probability that a sight line through a halo yields an absorber is
governed by two physical quantities: the mean covering fraction of
cold gas per halo and the fraction of halos that host cold gas
clouds. These two quantities are degenerate, and we parameterize their
product with the quantity $\cg$.

From equations (\ref{e.prho}) and (\ref{e.pW_prhob}), the probability
distribution function of equivalent widths is
\begin{equation}
\label{e.pw}
P(\wr|M) = \cg(M)\frac{2\,\pb(\wr|M)}{R_g^2}\frac{d\pb}{d\wr},
\end{equation}
where $\pb(\wr|M)$ is the inversion of equation
(\ref{e.w_rho2}), which is performed numerically. The derivative
of $\pb$ with respect to $\wr$ is
\begin{equation}
\label{e.dpdW}
\frac{d\pb}{d\wr} = \left(2A_W{\cal G}_0\right)^{-1} 
\left[\frac{\pb\times\tan^{-1}x}{(\pb^2 + a_h^2)^{3/2}} + 
\frac{1+x^2}{\sqrt{\pb^2 + a_h^2}} 
\frac{(a_h^2 + R_g^2)(\pb\times x)}{(R_g^2 - \pb^2)(a_h^2 + \pb^2)}\right]^{-1},
\end{equation}
where $x\equiv \sqrt{(R_g^2 - \pb^2)/(a_h^2 + \pb^2)}$.

Instead of parameterizing the mass dependence of $\cg$ using
power-laws or polynomials, we adopt a non-parametric approach and
specify the values of $\cg$ at four different masses and spline
interpolate between them. Thus, the form of $\cg$ is fully
non-parametric. The four mass values, $M_1$ though $M_4$ are $\log M_i
= 10.0$, $11.33$, $12.66$, and $14.0$. We denote the four parameters
of the $\cg$ function as $\kappa_1$, $\kappa_2$, $\kappa_3$, and
$\kappa_4$. We perform the interpolation in $\log M$ and $\log \cg$
and extrapolate to higher and lower masses.

In this smooth gaseous halo model, which we will refer to as the
``classical'' model, we adopt a double power law to account for the
mass dependence of the mean absorption per unit surface mass density
$A_W$, i.e.,
\begin{equation}
\label{e.A_W}
A_W(M) = \left\{ \begin{array}{ll}
	A_0\, M_{12}^{\aaa} & \hspace{1in}{\rm if\ \ } M_{12}\le 1 \\
	A_0\, M_{12}^{\aab} & \hspace{1in}{\rm if\ \ } M_{12}> 1. \\
	\end{array}
	\right.
\end{equation}
As a result, this classical model contains seven free parameters to be
constrained by the data.  Three govern the amount of absorption as a
function of halo mass, $A_0$, $\aaa$, $\aab$, and four characterize
the mass dependence of gas covering fraction $\kappa_1$-$\kappa_4$.

In summary, the gas density profile determines the relative absorption
between different impact parameters in a halo---the density profile
sets the {\it shape} of $P(\wr|M)$. The relative absorption between
halos of different masses is determined by $A_W$; at $s=0$, the line
of sight directly through the center of a halo, the absorption scales
as $\wr(0|M) \simeq M_{12}^{1/3}(A_W/10.9)$ where $A_W$ is in units of
($h$\,\AA\,cm$^2$/gm).  The incidence of the absorption is set by
$\cg$.  Some sightlines that pass through a halo will encounter no
absorbing gas, either due to the fact that the absorbing gas is highly
clumpy and the covering fraction is low or due to the fact that the
halo does not contain cold gas.
%The model
%must differentiate between scenarios in which all halos produce $\sim
%0.1$ \AA\ absorbers, and those in which $\sim 10\%$ of halos yield 1
%\AA\ absorbers. The latter may occur if the medium is highly clumpy,
%if the absorption is created by a mechanism such as an asymmetric
%outflow, or if only a subset of halos (e.g., star forming systems)
%contain extended cold gas halos. Separating $A_W$ and $\cg$ allows the
%model the freedom to distinguish between these scenarios.  There
%should be some physical connection between $A_W$ and $\cg$, since the
%amount of cold gas should relate to the covering fraction for a given
%set of cloud properties. But with the multiple parameters that
%determine $A_W$, that connection is lost in the degeneracies.

\begin{figure*}
\epsscale{1.0} 
\plotone{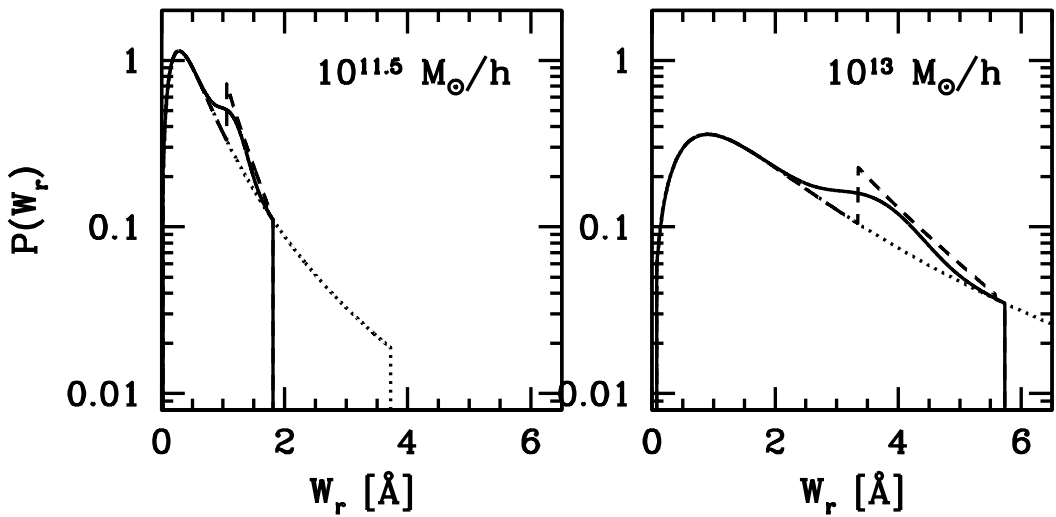}
\vspace{-7.5cm}
\caption{ \label{pw_curves} 
 The distribution of equivalent widths $P(\wr|M)$ for halos of
 $10^{11.5}$\hmsol\ (left panel) and $10^{13}$ \hmsol\ (right
 panel). In both panels, the dotted curve is $P(\wr|M)$ for a model in
 which the cold gas follows an isothermal density profile throughout
 the halo. The dashed curves represent at model in which the cold gas
 within $R=0.3\rg$ is reduced by a factor of $1-\fcold=0.95$. The
 sharp feature in the PDF is due slow change in $\wr$ as $\pb$
 approaches 0. The solid curve represents a model where this feature
 is smoothed out, as expressed in equation (\ref{e.pw_gaussian}).  }
\end{figure*}

%%%%%%%%%%%%%%%%%%
\subsection{Incorporating a Cold-Mode/Hot-Mode Transition}
%%%%%%%%%%%%%%%%%%

As we summarized in \S\ 1, the Mg\,II absorbers are known to originate
in primarily cool, photo-ionized gas. Recent numerical simulations
show that stable shocks develop as gas accretes onto a dark matter
halo, and above a critical mass threshold all gas is shock heated to
the virial temperature or above.  At low mass scales, however, gas
accreted onto halos is not shock heated, but remains cold as it
sinks to the center of the halo (e.g.,
\citealt{keres_etal:05}). \cite{dekel_birnboim:06} demonstrate that
this is due to the gas cooling time being too short with respect to
the compression time in low mass halos, thus shocks are not stable at
any radius within the halo.  

It is therefore expected that halos above a certain mass scale may not
contribute significantly to the observed Mg\,II absorbers due to shock
heating of their gas. In this second scenario, we include a
transitional mass scale, above which the gas becomes predominantly
shock-heated and unable to contribute to \mg2\ absorption. We denote
the shock radius $\rshock$. The absorption efficiency at $r<\rshock$
is reduced to a fraction $f_{\rm cold}$, which represents the fraction
of gas that is cold within the shock radius relative to gas outside the
shock. For impact parameters smaller than the shock radius, equation
(\ref{e.w_rho2}) is modified as the following,
\begin{equation}
\label{e.w_rho_shock}
\wr(\pb|M) = \wr(\pb|M)_{(\rshock=0)} - (1-f_{\rm cold})A_W\frac{2 {\cal G}_0}{\sqrt{\pb^2 + a_h^2}}\tan^{-1}
\sqrt{\frac{\rshock^2 - \pb^2}{\pb^2 + a_h^2}},
\end{equation}
where the first term on the left hand side is the result
from equation (\ref{e.w_rho2}), in which $\rshock=0$. Equation
(\ref{e.dpdW}) is calculated for $\pb<\rshock$ in the same manner.

Figure \ref{pw_curves} shows $P(\wr)$ (normalized to $\cg=1$) for
halos of $10^{11.5}$ \hmsol\ (left panel) and $10^{13}$ \hmsol\ (right
panel). The value of $A_W$ is set such that a sightline at an impact
parameter of $R_g/2$ through a $10^{12}$ \hmsol\ halo yields an
equivalent width of 1.5 \AA\ for comparison purpose.  In each panel,
we present three different probability distribution functions
(PDFs). The dotted curve shows equation (\ref{e.pw}), which represents
the $\wr$ distribution for a continuous gas density profile with no
shock. As the impact parameter approaches the edge of the halo, $\wr$
becomes small.  It is more likely to go through at halo at large
impact parameters, but the projected density drops steeply as $\pb$
approached $\rg$, thus $P(\wr)=0$ at $\wr=0$. The mode of the
distribution occurs at $\pb \sim 0.9\,R_g$, and at larger $\wr$ the
curve follows $\wr^{-3}$. The distribution reaches a maximum value of
$\wr\sim 3.6$ \AA\ at $\pb=0$.

The dashed curve shows the distribution of equivalent widths for a
halo with $\rshock = 0.3\,R_g$, and $\fcold=5\%$. The primary effect
of shock heated gas in the inner sphere is to reduce the maximum
$\wr$. In the case of a $10^{11.5}$ \hmsol\ halo, the largest
equivalent width is reduced by a factor of two.  For $\pb<\rshock$,
$P(\wr)$ has a very different form.  As $\pb$ falls just inside the
shock radius, $W$ decreases rapidly.  At $\pb\sim \rshock/2$,
$\wr(\pb)$ reaches a minimum value as the small fraction of cold gas
within $\rshock$ accumulates enough to contribute to the total
absorption.  The minimum value of $\wr$ depends on $\fcold$, but around
the minimum, the derivative $d\wr/d\pb$ becomes small, producing a
sharp feature in the total $P(\wr)$ curve at $\wr\sim 1$ \AA.  This
feature is even more pronounced in the PDF for the higher mass halo
(right panel), also with $\rshock = 0.3R_g$.
 
The sharp feature is due to the idealized nature of the calculation.
In practice, the propagation of the shocks is unlikely to be
spherically symmetric, and a halo of Poisson-distributed clouds will
also smear out this feature. To facilitate the calculations, however,
we approximate the distribution of $\wr(\pb<\rshock)$ as a Gaussian
centered on $\mu_W = \wr(\pb=\rshock/2)$ with width $\sigma_W =
[\wr(\rshock)/2 - \wr(0)]/2$. The amplitude of the Gaussian function
is set by the fraction of the cross section within the shock radius,
which for $\rshock = 0.3\,R_g$ is $9\%$. The Gaussian function is then
added to the PDF for $\pb > \rshock$. For this shocked halo model, the
probability distribution is \break
\begin{eqnarray}
\label{e.pw_gaussian}
P(\wr|M)  & =  & \cg(1-f_{\sigma})\frac{\pb(\wr|M)}{R_g^2}\frac{d\pb}{d\wr} + \nonumber \\
 & & \frac{\cg f_{\sigma}}{\sqrt{2\pi\sigma_W^2}}
\exp\left[\frac{-(\wr-\mu_W)^2}{2\sigma_W^2}\right],\,\,\, \wr\le W_{\rm max},
\end{eqnarray}
where $f_{\sigma} = \rshock^2/R_g^2$ is the probability that a sight
line has an impact parameter within $\rshock$, and $W_{\rm
  max}=max[\wr(\rshock),\wr(0)]$ from equation (\ref{e.w_rho}). We
note that $P(\wr)=0$ for $\wr>W_{\rm max}$. The results are shown in
the solid lines in both panels of Figure \ref{pw_curves}.  Instead of
a sharp peak in $P(\wr)$, the smooth model creates a `shoulder' in the
PDF, smoothing out the discontinuity in the dashed curve. The
parameters of the Guassian are chosen such that this shoulder tracks
the peak in the PDF as $\rshock$ changes.

In the one-dimensional simulations of \cite{birnboim_dekel:03} and
\cite{dekel_birnboim:07}, shock heating first become stable at the
inner regions of the halo, then propagates outward as the halo evolves
and becomes more massive. Thus at a given epoch, the width of the
transition region in their simulations is roughly a factor of $2-5$ in
halo mass.  In more realistic three-dimensional simulations, but in
which the shock fronts are not fully resolved, the fraction of cold
gas decreases linearly with $\log M$ over $\sim 1$ dex in mass
(\citealt{keres_etal:05, birnboim_etal:07}). We parameterize the mass
dependence of $\rshock$ as
\begin{equation}
\label{e.rshock}
\frac{\rshock}{\rg} = \rs0 + \ashock\log_{10} \m12,
\end{equation}
and restrict $\rshock/\rg$ to be no smaller then $0$ and no larger
than $1$.  We will refer to the ``transition scale'' as the mass at
which $\rshock/\rg =0.5$.  At low masses the shock radius is zero,
while in the transition region the fraction of the gas radius that is
within the shocked core increases linearly with $\log M$ until
$\rshock=\rg$. For halos with $\rshock=\rg$, the equivalent width is
calculated as
\begin{equation}
\wr(\pb|M) = f_{\rm cold}A_W\frac{2{\cal G}_0}{\sqrt{\pb^2 + a_h^2}}\tan^{-1}
\sqrt{\frac{\rg^2 - \pb^2}{\pb^2 + a_h^2}}.
\end{equation}

Nominally, this ``transition'' model has eight free parameters: $A_W$,
$\kappa_1$ through $\kappa_4$, $\rs0$, $\ashock$, and $\fcold$. In
this scenario, we leave $A_W$ to be invariant with halo mass and
attribute relevant mass-dependent gas fraction to $\rshock$ and
$\fcold$, in order to simplify the model analysis.  Removing freedom
from $A_W$ in favor of $\rshock$ is beneficial in that it highlights
the differences between the two scenarios and allows us to isolate
features from each model.  As we will show in \S\ 4, the classical
model and the transition model have distinct predictions for various
statistical quantities of the Mg\,II absorbers.  These distinctions
can be attributed to the different spatial distributions of the cold
gas in halos of different mass, and serve as strong constraints of the
models.

%%%%%%%%%%%%%%%%%%%%%%%%%%%%%%%%%%%%%%%%%%%%%%%%%%%%
% SECTION 3 SECTION 3 SECTION 3 SECTION 3 SECTION 3
%%%%%%%%%%%%%%%%%%%%%%%%%%%%%%%%%%%%%%%%%%%%%%%%%%%%

\section{Derivation of the Observed Statistical Properties of Mg\,II Absorbers}

Both the frequency distribution of Mg\,II absorbers as a function of
$W_r$ and their clustering amplitude with respect to LRG are known
from previous work.  Here we summarize these known statistical
properties from observations and derive theoretical formulae based on
the two scenarios described in \S\ 2.

%%%%%%%%%%%%%%%%%%%%%%%%%%%%%%%%%
\subsection{The Frequency Distribution Function}
%%%%%%%%%%%%%%%%%%%%%%%%%%%%%%%%%

The frequency distribution function is defined as the number of
absorbers per absorption equivalent width interval per unit redshift
path length.  Here we combine the results from \cite{prochter_etal:06}
for $\wr>1$\AA\ with the results from \cite{steidel_sargent:92} for
$0.3$\AA\ $\le \wr\le 1$\AA. The \cite{prochter_etal:06} data is
nearly complete ($> 95$\%) over the redshift range $0.4\le z\le 2.2$
for absorbers of $W_r>1$ \AA.  The \cite{steidel_sargent:92} data is
complete to $W_r=0.3$ \AA\ over the redshift range $0.2\le z\le 2.2$.

For empirical measurements, we take the raw number counts from
\cite{prochter_etal:06} and divide the number counts by the total
co-moving path length of all the lines of sight, $l_{\rm tot}$. The
total co-moving path length is calculated by integrating the
distance-redshift relation, weighted by the number of lines of sight
$g(z)$ at each redshift interval $dz$ as follows,
\begin{equation}
\label{e.comoving_path_length}
l_{\rm tot} = \int dz\,g(z)\left[ \om(1+z)^3 + \Omega_\Lambda\right]^{-1/2}.
\end{equation}
Equation (\ref{e.comoving_path_length}) yields the total co-moving
radial distance covered by the lines of sight to $\sim 45,000$ quasars
available in SDSS DR3. The total number of absorbers is $N_{\rm
abs}=4835$, with $l_{\rm tot}= 4.55\times 10^7$ \hmpc. There is no
systematic or cosmic variance error estimate for the frequency data,
so we apply Poisson fluctuations for estimating random errors,
including a $5\%$ `systematic' uncertainty for each of the 38 bins
that have at least one absorber. This additional systematic error only
alters the total uncertainty for the lowest few $\wr$ bins from the
Prochter et al.\ sample ($W_r=1-1.5$ \AA) and prevents those data from
driving the total $\chi^2$ when evaluating a model.  For $\wr\le 1$\AA,
we use the results from \cite{steidel_sargent:92}.  This is a much
smaller sample, but is valuable for constraining the shape of $d^2N/(dWdl)$.
The Steidel \& Sargent measurements are scaled to those of the
\cite{prochter_etal:06} data for $1$\AA $< \wr \le 1.3$\AA\ to account
for the difference in the total redshift survey paths of the two
samples.

To obtain theoretical predictions, we calculate the conditional
probability of a halo containing an absorber of a given strength
$P(\wr|M)$ as specified in \S\ 2 for a given set of parameters.  The
integral of the product of the cross-section weighted halo mass
function and the probability of detecting an absorber of $\wr$ in a
halo of a given mass leads to the frequency of the absorbers versus
$\wr$, i.e.\

\begin{equation}
\label{e.model_dndW}
\fr \equiv \frac{d^2N}{d\wr\,dl} = \int dM \frac{dn}{dM} \sigma_g(M) P(\wr|M),
\end{equation}

\noindent 
where $\sigma_g(M)=\pi \rg^2$ is the cross section of the gas halo and
$dn/dM$ is the halo mass function.  The frequency distribution in
equation (\ref{e.model_dndW}) has units of \AA$^{-1}\,($\hmpc$)^{-1}$
in co-moving coordinates.

We note that two scaling factors need to be accounted for before a
direct comparison between empirical measurements and model predictions
can be done.  First, the redshift distributions of the frequency data
and the bias data are different.  As discussed in \S\ 3.2 below, the
bias data are limited by the redshift range for LRGs at $0.35 < z <
0.8$.  We therefore must re-normalize the frequency function to the
effective redshift of the B06 sample at $z=0.6$.  The amplitude of the
frequency function rises between $z=0.5$ and $z=1$, and the ratio of
$dN/dz$ at $z=0.6$ to the full sample is 0.55.  Second, the amplitude
of $\fr$ based on a SDSS DR5 Mg\,II absorber sample is found to
increase by a factor of 1.1 (J.\ X.\ Prochaska, private
communication).  We have therefore included this correction in our
empirical data for comparisons with models\footnote{We note that the
  slope of the full \cite{prochter_etal:06} sample is $d\log N/dW =
  0.72$, a value consistent with the results from
  \cite{nestor_etal:05}, who calculate $d\log N/dW = 0.75$ over
  $0.4<z<0.8$ for a smaller sample culled from the SDSS early data
  release.}.

%%%%%%%%%%%%%%%%%%%%%%%
\subsection{The Mg\,II--LRG Cross-Correlation Function}
%%%%%%%%%%%%%%%%%%%%%%%

The cross-correlation function between absorbers and galaxies along
common lines of sight determines the clustering strength (and
therefore the relative bias) of the absorbers with respect to the
galaxies.  Given the mean mass scale of the halos that host the
galaxies, one can derive the mean mass scale of the halos hosting the
absorbers.  \cite{bouche_etal:06} presented the projected
cross-correlation function of Mg\,II absorbers and LRGs on scales of
$r_p=1-10$ \hmpc. The \mg2 -LRG sample covers a redshift range of
$0.35<z<0.8$. These authors calculate the relative bias of absorbers
to LRGs by taking the ratio of absorber-LRG cross-correlation function
to the LRG autocorrelation function,

\begin{equation}
\label{e.bhat}
\bhat \equiv \frac{b_W}{b_{G}} = \frac{\xi_{WG}}{\xi_{GG}}.
\end{equation}

\noindent Equation (\ref{e.bhat}) is only exact at large scales
because it assumes linear bias, but the scale dependence of halo bias
is largely independent of halo mass (\citealt{tinker_etal:05}) and
will cancel in the ratio.  We take the absolute bias of LRGs with
respect to dark matter to be $b_{\rm LRG}=1.85$ at $z=0.6$
(\citealt{white_etal:07}, who adopt the same cosmology as what is
assumed here), and estimate the absolute bias of the absorbers
following Equation (\ref{e.bhat}).

To compare the models to the data, we calculate the absolute bias of
absorbers and divide the calculation by $b_{LRG}$.  The bias data are
in four bins of $\wr$: [0.3,1.15] \AA, [1.15,2.0] \AA, [2.0,2.85] \AA,
and [2.85,4.0] \AA. The highest-$\wr$ bin has little influence on the
parameter constraints due to the small number of objects in that bin.
When comparing models to bias data, we take the number-weighted
average bias over the given range in $\wr$. The B06 sample is
incomplete for $\wr<1$ \AA.  The true frequency distribution
monotonically increases with smaller $\wr$, but the [0.3,1.15] \AA\
bin is dominated by $\sim 1$ \AA\ absorbers.  When calculating $\bb$
in this bin, we introduce the same incompleteness in the model by
multiplying $P(\wr|M)$ by $21.4\,(1-\wr)$ for $\wr<1$ \AA.  With this
factor, a model that accurately re-produces the true $d^2N/(d\wr\,dl)$
data will produce a frequency distribution that matches the shape of
the B06 sample.

In the context of the halo model, the bias of absorbers with respect
to the dark matter is equal to the mean halo bias, weighted by the
probability of finding an absorber $\wr$ at mass $M$;

\begin{equation}
\label{e.model_bias}
b_W = \frac{1}{\fr} \int dM \frac{dn}{dM} \sigma_g(M) b_h(M) P(\wr|M),
\end{equation}

\noindent where $b_h$ is the halo bias. For the halo mass function we
use the fitting function of \cite{warren_etal:06}, while for the halo
bias we use the fitting function of \cite{tinker_etal:05}. All halo
properties are calculated at the effective redshift of the B06
absorber sample. The redshift range of the data is large enough such
that integrating equations (\ref{e.model_dndW}) and
(\ref{e.model_bias}) over redshift will yield slightly different
results, but for the purpose of this study it is sufficient to choose
an effective redshift for all calculations. In our subsequent analyses,
the redshift evolution of $\fr$ will be used as an additional
constraint on the models.

\begin{figure*}
\epsscale{1.0} 
\plotone{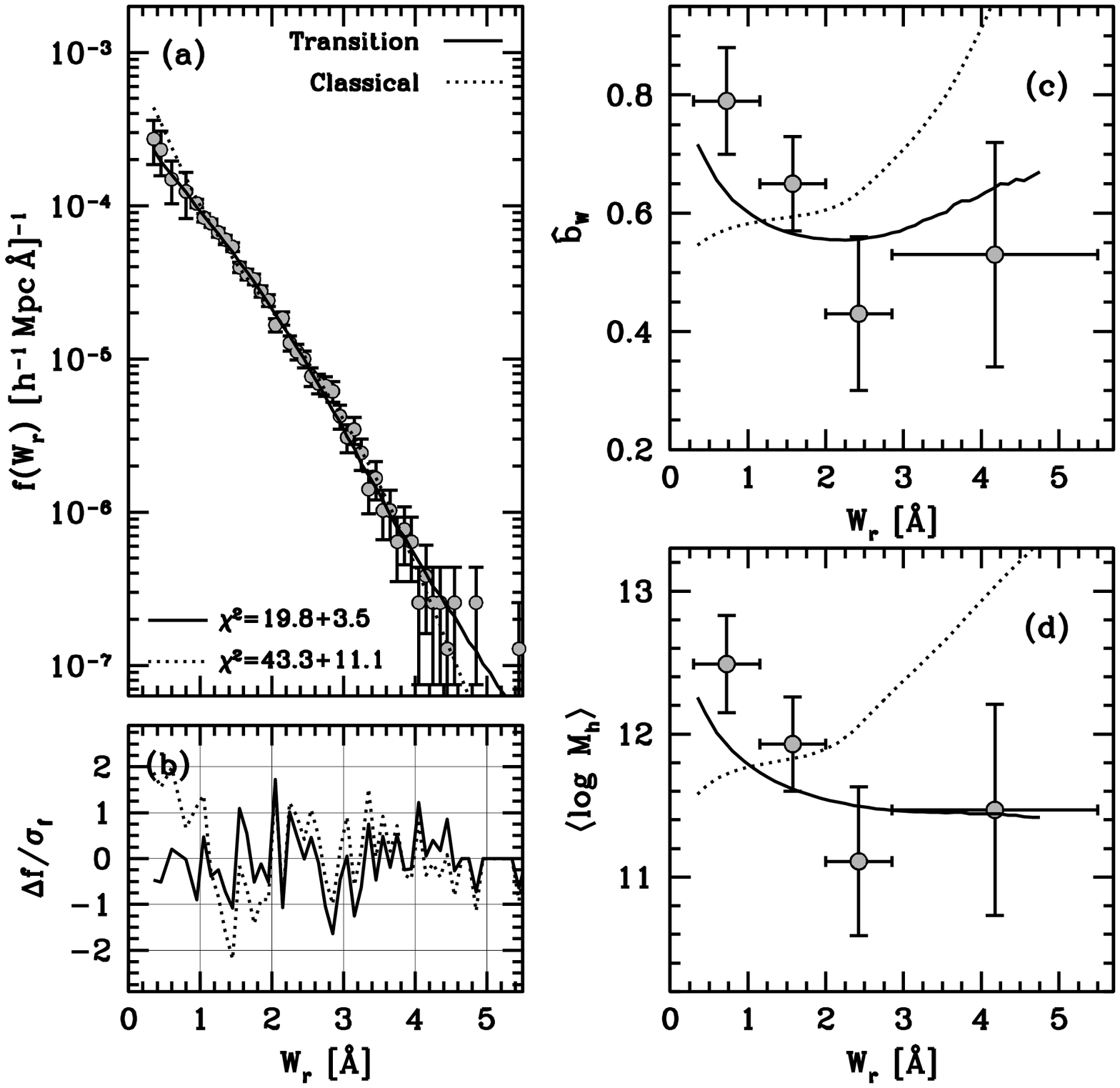}
%\vspace{-3cm}
\caption{ \label{stats} Panel (a) The frequency distribution of
  \mg2\ absorbers as a function of equivalent width $\wr$. The points
  with error bars represent a combination of the data sets of
  \cite{prochter_etal:06} ($\wr>1$\AA) and \cite{steidel_sargent:92}
  ($\wr<1$\AA). The transition between the two data sets is seen
  easily in the relative error bars between low and high $\wr$. The
  solid and dotted curves represent the best-fit models listed in
  Table 1. The solid curve is the model with a cold-hot transition in
  halo gas, while the dotted curve has a smoothly varying cold gas
  fraction as a function of halo mass. The $\chi^2$ values are broken
  into the $\chi^2$ for the frequency and bias data,
  respectively. Panel (b): The residuals of the model $\fr$ curves,
  normalized by the errors on the data. Panel (c): The bias of
  absorbers relative to LRGs, $\bb$, as a function of $\wr$. The
  points with errors are the measurements of
  \cite{bouche_etal:06}. The solid and dotted curves are the two
  models from panel (a). Panel (d): The mean logarithmic halo mass as
  a function of $\wr$. Points with error bars are the values inferred
  from the bias data by B06 (not used in the fit). Solid and dotted
  curves are the two best-fit models.}
\end{figure*}

%%%%%%%%%%%%%%%%%%%%%%%%%%%%%%%%%%%%%%%%%%%%%%%%%%%%%%%%%%%%%%%%%%%%%%%%
%  SECTION 4 SECTION 4 SECTION 4 SECTION 4 Results Section
%%%%%%%%%%%%%%%%%%%%%%%%%%%%%%%%%%%%%%%%%%%%%%%%%%%%%%%%%%%%%%%%%%%%%%%%

\section{Comparisons between Observations and Best-Fit Models}

The scenarios discussed in \S\ 2 have distinct predictions for the
empirical data discussed in \S\ 3.  We constrain the models using a
Monte Carlo Markov Chain technique that determines the best-fit
parameters based on comparisons of the data with random realizations
from various model predictions.  For each realization in the chain,
the total $\chi^2$ is the sum of the $\chi^2$ values from both the
frequency and bias data. We will refer to these quantities as $\xf$
and $\xb$, respectively. Because of the larger number of frequency
data and smaller errors on those data, $\xf$ dominates the total
$\chi^2$.  However, the $\xf$ values should be taken with caution,
because we have no true estimate of the cosmic variance.

%%%%%%%%%%%%%%%%%%%%%%%%%%%%%%%%%%%
% Table 1
%%%%%%%%%%%%%%%%%%%%%%%%%%%%%%%%%%%

%\begin{deluxetable}{ccccccccccccc}
\begin{deluxetable*}{ccccccccccccc}
\tablecolumns{13} \tablewidth{43pc} \tablecaption{Parameters of the
  Best-Fit Models} \tablehead{\colhead{Model} & \colhead{$A_W$} &
  \colhead{$\rs0$} & \colhead{$\ashock$} & \colhead{$\log \kappa_1$} &
  \colhead{$\log \kappa_2$} & \colhead{$\log \kappa_3$} & \colhead{$\log \kappa_4$} &
  \colhead{$\aaa$} & \colhead{$\aab$} & \colhead{$\fcold$} & \colhead{$\xf$} & \colhead{$\xb$} }

\startdata

Classical & $33.2$ & --- & --- & -1.721 & -0.012 & -0.198 & -1.763 & -0.172 & -0.176 & --- & 43.3 & 11.1 \\
Transition & $136$ & 1.02 & 1.03 & -9.27 & -0.205 & -0.006 & -0.168 & --- & --- & 0.061 & 19.8 & 3.5 \\

\enddata \tablecomments{ The units of $A_W$ are [$h$\,\AA\,cm$^2$/gm].} 

\end{deluxetable*}
%\end{deluxetable}

\begin{figure*}
\epsscale{1.0} 
\plotone{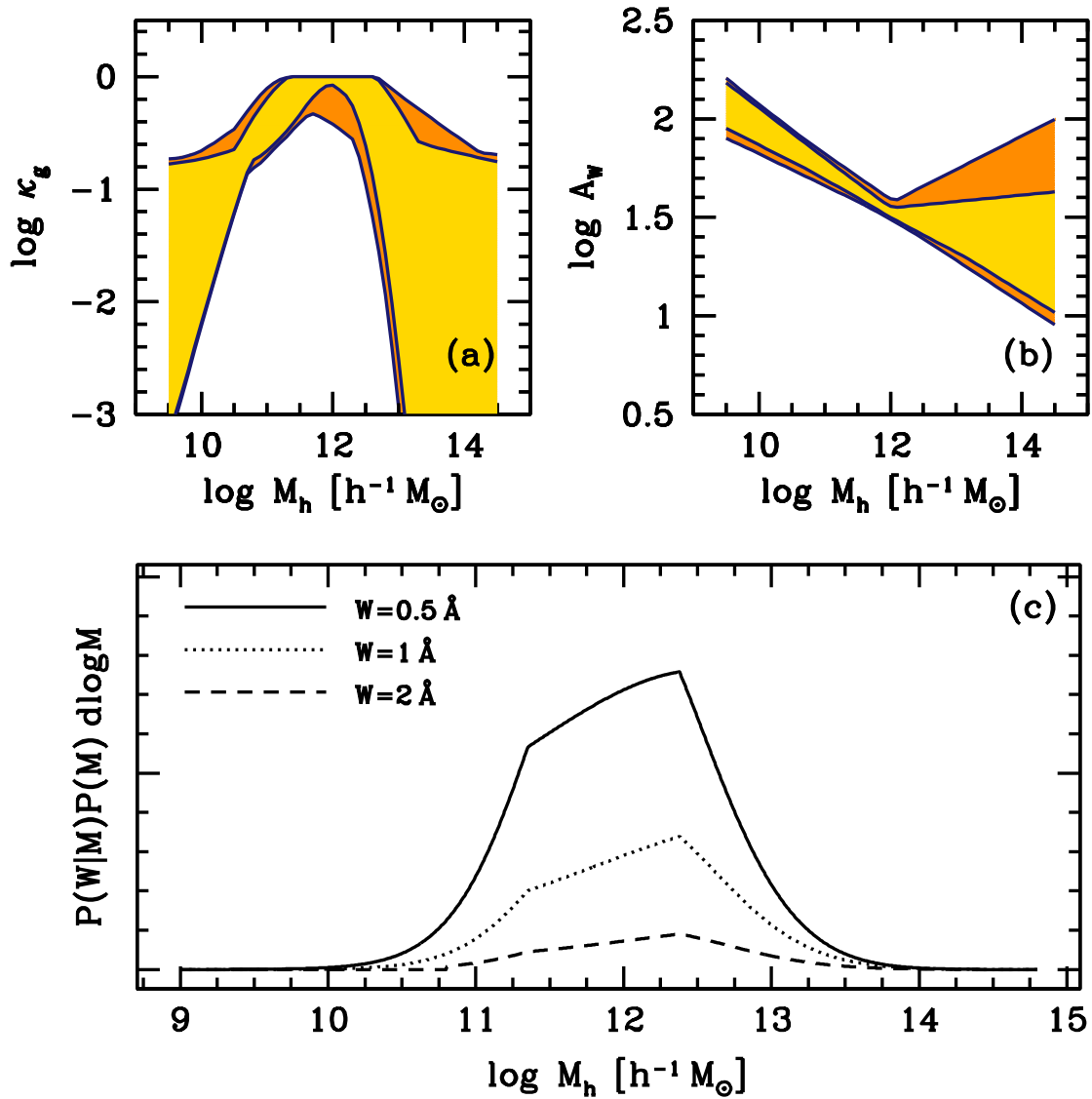}
%\vspace{-4.5cm}
\caption{ \label{all_137} Panel (a): Constraints on the covering
 fraction $\cg$ as a function of halo mass from the classical
 model. Inner and outer contours represent the range in models that
 produce $\Delta\chi^2<1$ and $4$ with respect to the best-fit
 model. Panel (b): Constraints on the mean absorption strength per
 unit gas mass density $A_W(M)$.  Contours are for the range of models
 described in panel (a). Panel (c): The occupation function of
 absorbers for three different equivalent width values: 0.5 \AA, 1
 \AA, and 2 \AA. The $y$-axis, in arbitrary units, is the probability
 that a sightline passes through a halo of mass $M$ ($P(M)$)
 multiplied by the probability that a halo of mass $M$ yields a $\wr$
 absorber ($P(\wr|M$)). }
\end{figure*}

\begin{figure*}
\epsscale{1.0} 
\plotone{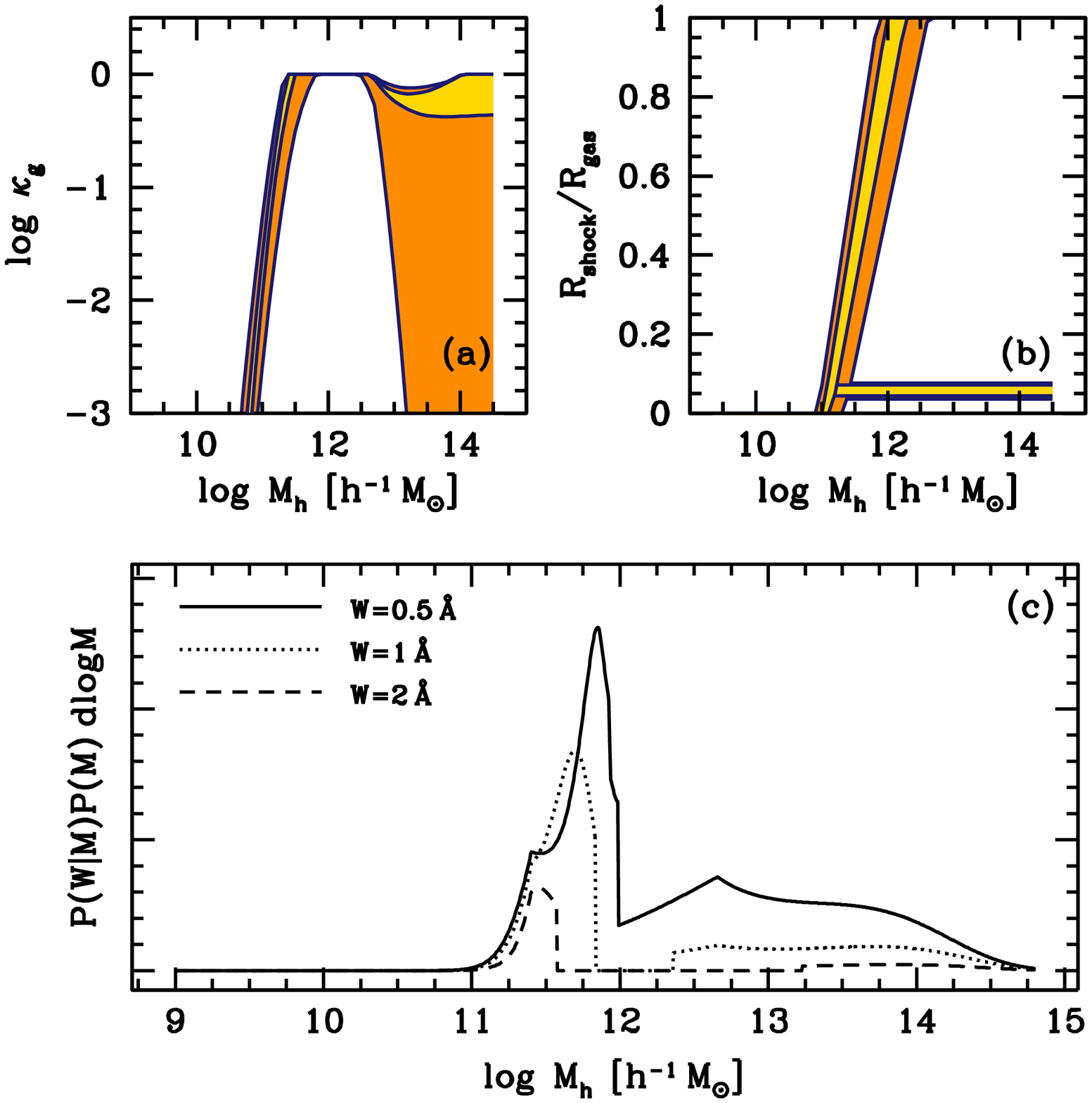}
%\vspace{-4.5cm}
\caption{ \label{all_136} Panel (a): Constraints on the covering
  fraction $\cg$ as a function of halo mass from the transition
  model. Inner and outer contours represent the range in models that
  produce $\Delta\chi^2<1$ and $4$ with respect to the best-fit
  model. Panel (b): Constraints on the shock radius $\rshock$
  (relative to the gas radius).  Contours are for the range of models
  described in panel (a). The horizontal bands to the right of the
  $\rshock$ contours represent the constraints on $\fcold$. Panel (c):
  The occupation function of absorbers from the best-fit model. The
  abrupt transition at $M=10^{12}$\hmsol\ is due to $\rshock$ becoming
  larger than $\rg$.}
\end{figure*}

%%%%%%%%%%%%%%%%%%%%%%%%%%%%%%%%%%%%%%
%\subsection{The Uniform Gaseous Halo Model}
\subsection{The Classical Gaseous Halo Model}
%%%%%%%%%%%%%%%%%%%%%%%%%%%%%%%%%%%%%%

As described in \S\ 2.2, the classical model contains seven free
parameters, including three that characterize the absorbing gas
column, $A_W$, for halos of different mass scales and four that
characterize gas covering fraction $\kappa_1$-$\kappa_4$ of individual
halos.  The mass dependence in $A_W$ allows the predicted $\wr$ versus
impact parameter $\pb$ relation (Equation \ref{e.w_rho2}) to vary
smoothly with halo mass.  The best-fit parameters are listed in Table
1.  Figure \ref{stats} compares observational data with predictions
based on the ``classical'' model (dotted curves).

We present in panel (a) the frequency data and in panel (b) the
residuals normalized by the corresponding error of each bin.  The
frequency data are adequately fit by this model, yielding $\xf=43.3$
for 41 data points and seven free parameters.  The classical model
matches the abundance of strong absorbers, $\wr>2$ \AA, but at lower
equivalent widths the residuals become large.  Specifically, this
model overproduces the number of absorbers by nearly 60\% at $W \sim
0.3$ \AA\ due to a large contribution from low-mass halos.

We present the comparison for the halo bias of the absorbers in panel
(c) and the inferred mean halo mass in panel (d).  The bias data
(points with errorbars) are derived from Mg\,II--LRG cross-correlation
measurements.  The relative bias $\bhat$ derived from observations
decreases from 0.8 at $\langle \wr \rangle \approx 0.8$ \AA\ to nearly
0.4 at $\langle \wr\rangle \approx 2.4$ \AA.  The measured bias for
$\wr\ge 2.8$ \AA\ increases slightly to 0.5, but the number of objects
in this bin is small and it has little influence on our fitting
results.  Using a simple mean-mass approximation, B06 infer a decrease
in mean halo mass of nearly 1.5 dex over this range in $\wr$.  The
best-fit classical model is unable to reproduce the inverse
correlation between $\bhat$ and $\wr$.  As expected, the bias
monotonically increases with $\wr$, placing the weakest absorbers in
$M=10^{11.5}$ \hmsol\ halos while locating the strongest absorbers in
the nearly same halos as LRGs.

Figures \ref{all_137}a-\ref{all_137}b present the parameter
constraints for this model.  In both panels, the inner and outer
shaded regions show the allowed parameter range, constrained
respectively by $\Delta\chi^2<1$ and $<4$ from the best-fit
model. Figure \ref{all_137}a presents the constraints on the mean gas
covering fraction $\cg$ as a function of halo mass. The function peaks
at $\log M\approx 12$; nearly all halos of $\log M\approx 11.5-12.5$
are expected to produce an absorber if the impact parameter is less
than $\rg$ to a background quasar. At lower and higher masses, $\cg$
rapidly falls off, but the constraints become weak at the extrema of
the mass range. Given the constraints, there is little contribution to
\mg2\ absorption from halos more massive than $10^{13}$\hmsol\ and
less massive than $3\times 10^{10}$\hmsol. 

Figure \ref{all_137}b shows constraints for the absorption column
$A_W(M)$.  For low-mass halos, the constraints on the slope and
normalization are strong; $10^{12}$\hmsol\ halos produce $\sim 1$\AA\
absorbers. The best-fit model has a single power-law, with $\aaa
\approx \aab \approx -0.17$. The negative values of $\aaa$ and $\aab$
increase the absorption efficiency in lower mass halos, but not
sufficiently to produce an anti-correlation between $\wr$ and
$M$. A slope more negative than $-1/3$ in $A_W(M)$ versus $M$ is
required to make the mean $\wr$ constant as a function of $M$, but
these negative values would violate the constraints on the frequency
distribution by overproducing high-$\wr$ absorbers.

At $M>10^{12}$ \hmsol, the constraints on $A_W$ are poor because
$\cg$ falls off rapidly at these masses. There is a degeneracy between
$\kappa_4$ and $\aab$; if $\cg$ falls off for $M>10^{12}$\hmsol,
high-$\wr$ systems can only be produced if $\aab$ is large. However,
this steepens the slope of $\bb$ and these models fare even worse in
comparison to the bias data. The models which produce the lowest $\xb$
have negative values of both $\aaa$ and $\aab$.

Figure \ref{all_137}c shows the relative contribution to absorber
counts from halos of different mass based on the best-fit model.  We
refer to this mass distribution as the ``occupation function''.  The
curves show three values of $\wr$: 0.5 \AA, 1 \AA, and 2 \AA, and the
relative areas under the curves represent the relative number of
absorbers. The $y$-axis, in arbitrary units, is the probability of
detecting an absorber of $\wr$ in a halo of mass $M$, i.e.,
$P(\wr|M)P(M)$, where $P(M)$ is the probability that a given line of
sight intersects a halo of mass $M$ based on a cross-section weighted
halo mass function, $\sigma_g(M)\,dn/dM$.  These curves have the
Gaussian-type shape of the $\cg$ function, with the sharp edges near
the peak resulting from the mass range over which $\cg$ is truncated
at unity.  The mass dispersion is $\sim 1$ dex in $\log M$ for all
three curves.  The modes of each curve move to higher mass, producing
the increasing trend of $\wr$ with $M$ shown in Figure \ref{stats}d.

In summary, the ``classical'' model with a smooth gaseous halo {\it cannot}
fit both the frequency data and the bias data simultaneously.  Under
this scenario, more massive halos are expected to contain more
absorbing gas, resulting in a positive correlation between $\wr$ and
halo mass $M$.  Forcing the mean absorption $A_W$ per halo to decline
steeply with increasing halo mass would improve the model fit to the
bias data, but at the same time it would over-produce the abundance of
strong ($\wr>2$ \AA) absorbers.

%%%%%%%%%%%%%%%%%%%%%%%%%%%%%%%%%%%%%
\subsection{The Significance of Shock-Heated Gas in Massive Halos}
%%%%%%%%%%%%%%%%%%%%%%%%%%%%%%%%%%%%%

The best-fit parameters of the transition model are also listed in
Table 1.  This transition model has eight free parameters: $A_W$,
$\kappa_1$ through $\kappa_4$, $\rs0$, $\ashock$, and $\fcold$.  As
discussed in \S\ 2.3, we have left $A_W$ to be invariant with halo
mass and attributed relevant mass-dependent gas fraction to $\rshock$
and $\fcold$.  Comparisons between observations and predictions based
on the transition model are also presented in Figure \ref{stats}
(solid curves).

The addition of the cold-hot transition produces a best-fit model that
precisely fits the frequency data over the entire $W_r$ range,
including those points at $W<1$\AA, with $\xf=19.8$.  The model also
reproduces the anti-correlation between $\wr$ and $\bhat$ for $\wr<2.5$
\AA, shown in Figure \ref{stats}c.  Although the slope of the trend is
not as strong as that seen in the B06 measurements, $\xb=3.5$ for four
data points. The relative bias in the model decreases from $\bhat=0.7$
at $\wr=0.4$\AA\ to $\bhat=0.55$ for $\wr=2$\AA.  This declining trend
in the absorber bias is due to a decrease in the mean halo mass with
increasing $\wr$ (Figure \ref{stats}d). The mean mass falls by nearly
a decade from $\wr=0.4$\AA\ to $\wr=4$\AA.\footnote{We note that the
  masses derived by B06 depend on both the model used and the
  cosmology adopted, and they are not utilized in the fit. Even a
  perfect fit to the bias data in panel (c) will not necessarily yield
  the B06 masses in panel (d).}

The total $\chi^2$ for this model is 23.3 for 37 degrees of
freedom. In a purely statistical sense this suggests that the model
has too many free parameters. However, we note once again that the
errors on $\fr$ are estimates only, and we wish to include as much
freedom in the model to prevent the frequency data from driving the
constraints on the halo occupation.  In this sense, $\xb$ is the most
relevant quantity, noting that eight free parameters is barely
sufficient to reproduce the bias measurements.

Figures \ref{all_136}a and \ref{all_136}b present the parameter
constraints for the transition model. The covering fraction is
narrower and sharper than the classical model: $\cg=1$ for halos of
$\log M\approx 11.5-12.5$ and cuts off sharply at lower masses. Over
nearly a decade in halo mass, all models with $\Delta\chi^2<4$ yield
$\cg=1$. However, the transition from cold mode ($\rshock/\rg=0$) to
hot mode ($\rshock/\rg=1$) begins at $M\simeq 10^{11}$ \hmsol\ and ends at
$M\simeq 10^{12}$ \hmsol.  This indicates that massive halos of $\log M\ge
12$ contribute to the observed Mg\,II statistics through the presence
of cold flow in the shock-heated halos. The constraints on the cold
fraction within the shock are strong, with the best-fit model yielding
$\fcold=0.06$ with an allowed range of $\pm 0.03$ for models with
$\Delta\chi^2<4$.  Both the transitional mass scale and cold gas
fraction in massive halos agree well with the simulations of
\cite{keres_etal:05} and \cite{birnboim_etal:07}. In the post-shock
regime $\log M>12.5$, the constraints on the covering factor are poor;
models with a low $\kappa_4$ yield better fits to the frequency data
while high covering factors at $10^{14}$ \hmsol\ produce higher bias
factors and lower $\xb$.

Figure \ref{all_136}c shows the occupation functions for the best-fit
model.  For halos of $M = 10^{11}-10^{12}$\hmsol, the absorber counts
rises as $\cg$ increases with halo mass as shown in panel
(a). For $\wr=0.5$ \AA, the occupation function peaks are $10^{12}$
\hmsol, then abruptly drops when $\rshock = \rg$ and all halos enter
the hot phase. For $\wr=1$ and 2 \AA, the peak in the occupation
functions occur at $10^{11.7}$ \hmsol\ and $10^{11.5}$ \hmsol,
respectively.  This trend is due to the inside-out nature of the
heating; higher-$\wr$ systems are produced at smaller impact
parameters and the shock radius envelopes that radius at lower halo
masses. Above $10^{12}$ \hmsol, the occupation has a tail out to
$10^{14}$ \hmsol. In the best-fit model, $\kappa_4 = 0.67$, implying
that a significant fraction of cluster-sized halos contain some cold
gas with a high covering fraction. However, we note again that the
constraint on $\kappa_4$ is poor due to the low frequency of such
massive halos.

For $\wr\ge 1$ \AA\ absorbers, halos in the post-shock regime do not
initially have enough gas to produce absorbers of this strength, and
so the distribution of halo masses occupied by these absorbers becomes
bimodal; only at $M>10^{12.5-13}$\hmsol\ is there enough cold gas to
produce absorbers of this strength. For $\wr=2$ \AA, for example,
halos between $10^{11.6}$ and $10^{13.2}$ \hmsol\ halos do not produce
strong absorbers. This bimodal distribution is what causes $\bhat$ to
increase for $\wr>3$ \AA, while $\logmbar$ remains constant in Figure
\ref{stats}; bias increases non-linearly with mass, weighting
high-mass halos more in $\bhat$ than in $\logmbar$.

\begin{figure*}
\epsscale{1.0} 
\plotone{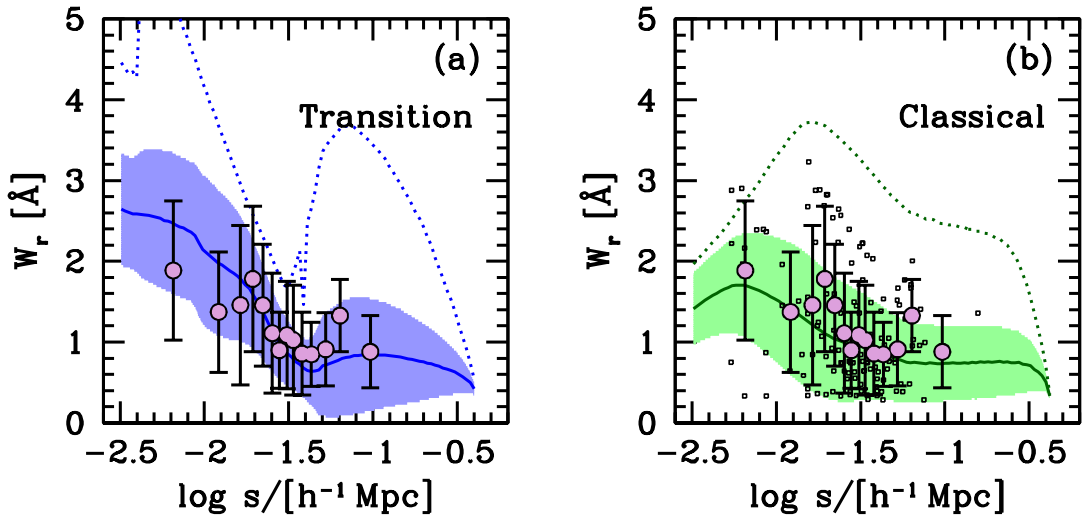}
\vspace{-7.5cm}
\caption{ \label{rhob} The distribution of equivalent width as a
  function of impact parameter. In both panels, the points represent
  the mean $\wr$ as a function of $\log\pb$ from a combination of four
  sets of data, while the errors show the dispersion about the
  mean. Panel (a) shows the predictions for the transition model. The
  solid curve plots the mean $\wr$ while the shaded region represents
  the predicted dispersion. The dotted curve is the upper-99\% bound on
  the distribution of $\wr$ at each $\log\pb$. Panel (b) presents the
  predictions for the classical model. The small open squares
  represent each individual system in the observational sample. The
  $x$-axis is in physical units. }
\end{figure*}

In summary, the transition model can simultaneously re-produce the
observed frequency data and bias data of Mg\,II absorbers.  The
anti-correlation between $\wr$ and $M$ in the transition model is
produced through two mechanisms. First, as the shock propagates
outward from the center of a massive halo, high-$\wr$ systems are no
longer produced, but low-$\wr$ systems are produced {\it more
  frequently} because sight lines at low impact parameters
($\pb\le\rshock$) now produce weaker absorbers. This is demonstrated
in Figure \ref{all_136}c by the location of the cutoff in the
occupation function---as $\wr$ increases the location of the cutoff
moves to lower masses. Second, the small amount of cold gas in the
post-shock regime (masses at which $\rshock=\rgas$) yields mostly
low-$\wr$ absorbers in high mass halos. An instantaneous transition
between the cold-mode and hot-mode with a non-zero cold fraction could
produce a weak $\wr-M$ anti-correlation. A broad transition region (a
small value of $\ashock$) with $\fcold=0$ could also produce a weak
anti-correlation. In this model the two effects work together to fit
the measurements.

%%%%%%%%%%%%%%%%%%%%%%%%%%%%%%%%%%%%%%%%%%%%%%%%%%%%%%%%%%%%%%%%%%%%%%%%
% SECTION 5 SECTION 5 SECTION 5 SECTION 5 SECTION 5 SECTION 5 
%%%%%%%%%%%%%%%%%%%%%%%%%%%%%%%%%%%%%%%%%%%%%%%%%%%%%%%%%%%%%%%%%%%%%%%%
\section{Additional Observables Predicted by the Models}

We have demonstrated that the transition model can re-produce both the
observed frequency distribution function and the Mg\,II--LRG relative
bias data.  Given the known $W_r(s|M)$ constrained by the data in \S\
4, and known mass distribution function of the dark matter halos, we
can also predict (1) the correlation between $\wr$ and galaxy impact
parameter $s$, and (2) the impact parameter distribution of Mg\,II
absorbing galaxies $P(s)$.  Here we derive these properties on the
basis of the gaseous halo model described in \S\ 2 and the best-fit
parameters constrained by the frequency and bias data in Table 1.

%%%%%%%%%%%%%%%%%%%%%%%%%%%%%%%%%%%%%%%%%%%%%%%%%%%%%%%%%%%%%%%%%%%%%%%%
\subsection{The Absorber $\wr$ vs. Galaxy Impact Parameter Correlation}
%%%%%%%%%%%%%%%%%%%%%%%%%%%%%%%%%%%%%%%%%%%%%%%%%%%%%%%%%%%%%%%%%%%%%%%%

\begin{figure}
%\epsscale{1.0} 
\epsscale{2.3} 
\plotone{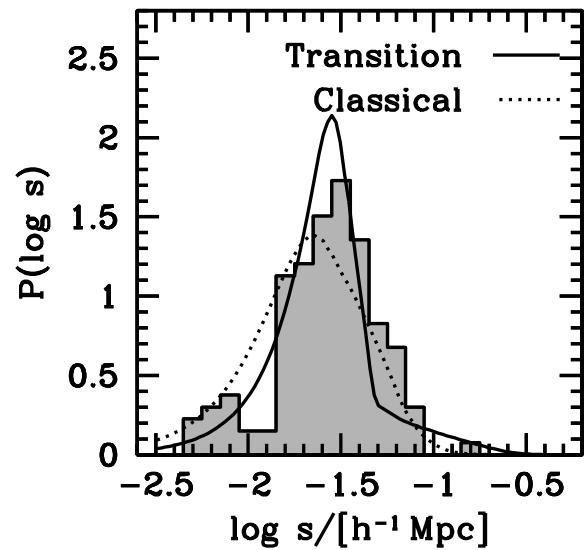}
\vspace{-8cm}
\caption{ \label{drho} The distribution of impact parameters for all
  absorbers with $\wr>0.3$ \AA. The shaded histogram represents the
  data presented in Figure \ref{rhob}. The solid and dotted curves show
  the predictions from the transition and classical models,
  respectively. The $x$-axis is in physical units. }
\end{figure}

To estimate the correlation between $\wr$ and impact parameter, as
well as the distribution of impact parameters, we adopt the center of
each halo as the location of the corresponding absorbing galaxy.  At a
given impact parameter, the distribution of equivalent widths is then
determined by

\begin{eqnarray}
\label{e.dW_rhob}
P(\wr|\pb) & = & \int dM \frac{dn}{dM} \sigma_g(M)\,P(\wr|M,\pb)\,P(\pb|M) \\
           & = & \int dM \frac{dn}{dM} \sigma_g(M)\,\delta(\wr(\pb|M)-\wr)\,\kappa_g(M)\,P(\pb|M) \nonumber
\end{eqnarray}
where $P(\pb|M)$ is the probability distribution of impact parameters
of random sightlines defined in equation (\ref{e.prho}) and
$\delta(x)$ is a delta function. From equation (\ref{e.dW_rhob}), the
total number of absorbers at a given impact parameter---and thus the
distribution of impact parameters as a function of $\pb$---is obtained
be integrating $P(\wr|\pb)$ over all $\wr$.

To use galaxy-absorber pair statistics to constrain the model
parameters requires a Mg\,II absorber survey carried out along the
sightlines where the galaxies positions are known {\it a priori}.
Typical absorbing galaxy studies focus on searching for galaxies that
give rise to known absorbers.  These galaxy-absorber pairs collected
from these studies are not suitable for constraining the covering
fraction of Mg\,II gas random halos because sightlines that do not
encounter a Mg\,II absorber are by design excluded from the sample.
Therefore, the impact parameter distribution established based on
these {\it absorber oriented} studies is expected to be skewed.  A
galaxy--Mg\,II absorber pair sample established in a {\it galaxy
  oriented} survey is only mentioned in Tripp \& Bowen (2005), but the
sample is small and details are not available in the literature.

In the subsequent discussion, we have compiled an {\it ad hoc} sample
of 133 galaxy--Mg\,II absorber pairs from the literature
(\citealt{steidel:95}; \citealt{kacprzak_etal:07};
\citealt{bergeron_boisse:91}; B06), in which the galaxies are found
for known Mg\,II absorbers. The mean redshift of these samples is
$z\sim 0.6$.  We note a potential bias in this heterogeneous pair
sample.  Because of different selection criteria, these pairs are most
likely incomplete at impact parameters $s\lesssim 15$ \hkpc\ due to
blending of the galaxy light with the QSO.  The likelihood of
mis-identifying the associated galaxy at $s\gtrsim 60$ \hkpc\ is
non-negligible, if the true associated galaxy is too faint to be
detected.  While we cannot address the covering fraction of Mg\,II gas
in typical galaxies, this pair sample offers a test of our model based
on the observed $\wr$-$\log\pb$ distribution of known absorbers. This
sample also offers an important consistency check on our choice of gas
radius.

In Figures \ref{rhob}a and \ref{rhob}b, we compare the predicted
$\wr$-$\log\pb$ relations from the classical and shocked halo models
to observational data.  Each point with error bars in Figure
\ref{rhob} is the mean and dispersion from every 10 galaxy-absorber
pairs, ordered by $\log\pb$. It should be noted that the error bars do
not represent the error in the mean but rather the variance in $\wr$
at fixed $\pb$, which is a physically meaningful quantity that can be
compared to models as well. The data show a weak trend of decreasing
$\wr$ with larger impact parameter, with $\wr \sim 1.6$ \AA\ at $-2.3
< \log \pb <-1.6$ and $\wr \sim 0.9$ \AA\ at $\log\pb\gtrsim 1.6$. The
dispersion in $\wr$ also decreases from low to high $\pb$. There is an
``upper envelope'' to the data in the $\wr$-$\log\pb$ plane such that
the highest-$\wr$ systems are preferentially at lower impact
parameters (see panel $b$). All distances are in physical units.
 
In Figure \ref{rhob}a, the predictions for the shocked halo model are
shown. The solid curve is the mean equivalent width $\wbar$ as a
function of $\log\pb$. The shaded area is the predicted dispersion in
$\wr$. The size of the dispersion reflects the distribution of halo
masses that contribute $\wr>0.3$ \AA\ absorbers. The dotted curve is
the upper-99\% bound on the distribution of $P(\wr|\pb)$. For
$\log\pb>-1.8$, both the predicted $\wbar$-$\log\pb$ mean relation and
the associated dispersion in $\wr$ are well rendered in this model.
At fixed $\pb$, the mean $\wr$ and the dispersion around that value
reflect the mean and dispersion in halo mass. A broader distribution
of halo masses in $\cg$ will produce a larger dispersion in $\wr$ at
fixed impact parameter. {\it The decrease in $\wbar$ with $\log\pb$ results
from the fact that in the models the gaseous halos are being probed
preferentially at their edges at larger $\pb$, where the gas density
is low.} The dispersion decreases because the range in halo masses that
can be probed decreases; at large $\pb$, smaller halos can no longer
be detected and the variance in $\wr$ decreases. The inflection in the
model predictions at $\log\pb \sim -1.5$ occurs at the cold-hot
transition scale, and only halos in the hot mode are large enough to
produce impact parameters above 50 \hkpc. In the immediate post-shock
regime, only very weak absorbers can be produced, so the dispersion in
$\wr$ reaches a local minimum at this scale.

For comparison, the predictions from the classical model are shown in
Figure \ref{rhob}b. The open squares in this figure show the
individual absorbers, elucidating the upper envelope of $\wr$ in the
data.  At $\log\pb>-1.8$, the $\wbar$-$\log\pb$ trend is similar to
the transition model, and the upper-99\% bound matches the
observational sample.  B06 argue that the upper envelope and the trend
of $\wbar$ in the data necessitate a $M$-$\wr$ anti-correlation.  The
results of the classical model invalidate this argument; stronger
absorbers occur at lower $\pb$ simply because, at fixed halo mass,
$\wr(\pb)$ monotonically increases toward the center of the halo.

\subsection{The Impact Parameter Distribution of Mg\,II Galaxies}

Figure \ref{drho} shows the distribution of impact parameters for
absorbers of all strength, in comparison to model predictions. The
gray shaded histogram established from the heterogeneous pair sample
exhibits some hint of incompleteness of pairs at $\log\pb<-1.8$, while
the distribution at larger $\log\pb$ appears relatively well
sampled. The distribution of empirical data is roughly Gaussian in
$\log\pb$ with slight negative skewness. This asymmetry may be
physical, but could also be due to incompleteness. The solid curve,
showing the model prediction, locates the mode at 30\hkpc, in good
agreement with the data. The shape and dispersion are also in good
agreement with the data. The distribution predicted by the classical
model places the mode at roughly $\sim 20$ \hkpc, and the distribution
is broader than both the transition model and the data, reflecting the
broad shape of the $\cg$ function in the best-fit classical model.

In a broader sense, the results of Figure \ref{drho} support our
choice of $\rgas\approx R_{200}/3$ for the extent of the \mg2\ gas. If
$\rgas=R_{200}$, the modes of each histogram would be shifted by 0.5
dex, making them incompatible with the data. The predictions in
Figures \ref{rhob}a and \ref{rhob}b would also be shifted over by the
same amount. Setting $\rg=R_{200}$ would have no affect on the
accuracy of the fits to the frequency or bias data in Figure
\ref{stats}. Increasing the total gas cross section $\sigma_g(M)$ by a
factor of 9 can be compensated for by lowering $\cg$ by the same
factor. But values of $\cg$ that low would conflict with observational
estimates of the covering factor, which vary between 50\% and 100\%
(\citealt{steidel_etal:97, tripp_bowen:05}). This demonstrates the
potential of a complete sample of absorber-galaxy pairs for
constraining models. Finally, we note that a representative sample of
Mg\,II absorbing galaxies will also allow us to derive more
quantitative constraints on the luminosity distribution of the
absorbing galaxy population.

%%%%%%%%%%%%%%%%%%%%%%%%%%%%%%%%%%%%%%%%%%%%%%%%%%%%%%%%%%%%%%%%%%%%%%%%
%  DISCUSSION SECTION 6 DISCUSSION SECTION 6 DISCUSSION SECTION 6
%%%%%%%%%%%%%%%%%%%%%%%%%%%%%%%%%%%%%%%%%%%%%%%%%%%%%%%%%%%%%%%%%%%%%%%%

\section{Summary and Discussion}

We have presented a theoretical approach for interpreting the
statistical properties of Mg\,II absorbers based on the halo
occupation framework. The underlying assumption of the model is that
all absorbers originate in the dark matter halos that host galaxies,
and that the strength of an absorber is proportional to the amount of
cold gas projected along the line of sight though the halo.  The two
quantities that we specify in the model are (1) the mean absorption
strength per unit surface gas mass density $A_W(M)$, and (2) the mean
covering factor of the cold gas $\cg(M)$.  Both of these quantities
are allowed to vary as a function of halo mass and together they
determine the conditional probability distribution of $\wr$ as a
function of halo mass, $P(\wr|M)$. 

For the conditional probability distribution function of equivalent
widths, $P(\wr|M)$, the first quantity $A_W$ determines how broad the
distribution is, the second quantity $\cg$ governs the normalization
of the PDF.  In this statistical approach, we parameterize the physics
that governs the gaseous halo and then allow the observational data to
determine the best-fit values of the model. In this way, our approach
is complementary to ab initio models for QSO absorption systems. The
halo occupation of cold gas driven by the data can be compared to both
analytic models and hydrodynamical simulations of cosmological
structure growth.

We find that models in which the cold gas fraction varies smoothly
with halo mass are not able to properly match either the frequency of
absorbers or the observed anti-correlation between $\wr$ and mean halo
bias. Models that incorporate a rapid transition in the cold gas
fraction---from low-mass halos that contain predominantly cold gas to
high-mass halos that contain predominantly shock-heated gas and cannot
contribute significantly to the observed Mg$^+$
absorption---accurately fit the observational data.

This latter model is consistent with the results of recent numerical
simulations that display a revised picture in the mass assembly of
galaxies (\citealt{birnboim_dekel:03, keres_etal:05,
dekel_birnboim:06, birnboim_etal:07}).  The classical paradigm of
galaxy formation states that all gas accreted onto the halo is shock
heated at the virial radius and cooling begins at the inner regions of
the halo. In the revised picture, shock heating does not occur in
halos below a transitional mass scale, where the compression time is
longer than the cooling time.  Beyond the transitional mass, however, 
shock heating develops and becomes stable.  In the revised scenario, 
gas {\it heating} occurs as an inside-out process until the entire
halo is within the shock radius.  But in high-mass halos that is
dominated by a ``hot-mode'' accretion, some fraction of cold streams
can still penetrate through the shock heated gas and reach the center
of these halos.

In our analysis, we find a best-fit transitional mass scale of
$M=10^{11.5}$ \hmsol, in excellent agreement with predictions from
these numerical simulations.  In addition, we constrain the cold gas
fraction in high-mass halos to be $\sim 6\%$ relative to the cold gas
fraction in the pre-shock regime.  The best-fit model predicts that
the majority of absorbers with $\wr\gtrsim 2$ \AA\ arise in the
pre-shock mass regime, while lower-$\wr$ systems are equally
predominantly contributed by post-shock halos.  In the following
section, we discuss in more detail of the implications of these
constraints in the transition halo model.

%%%%%%%%%%%%%%%%%%%%%%%%%
% FIGURE
%%%%%%%%%%%%%%%%%%%%%%%%%
\begin{figure}
\epsscale{2.2} 
%\epsscale{1.0} 
\plotone{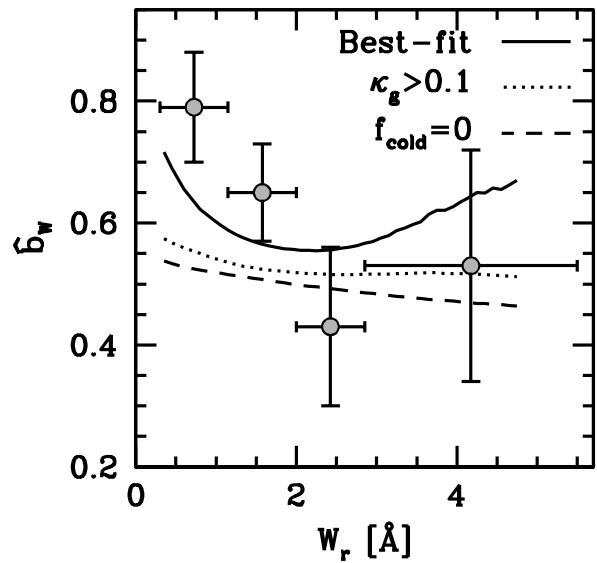}
\vspace{-8cm}
\caption{ \label{constraints} Relative bias $\bhat$ as a function of
  $\wr$, demonstrating the significance of the constraints on the mean
  gas covering fraction at low masses and the cold gas fraction at
  high masses. The points are the B06 data, and the solid curve is the
  best-fit transition model from Figure \ref{stats} and Table 1. The
  dotted curve is the best-fit model adjusted such that $M<10^{11}$
  \hmsol\ halos have $\cg=0.1$. The dashed curve is the best-fit model
  with $\fcold$ set to zero. }
\end{figure}
%%%%%%%%%%%%%%%%%%%%%%%%%
% FIGURE
%%%%%%%%%%%%%%%%%%%%%%%%%

\subsection{Covering Fraction of Cold Gas, $\cg$}

The four-point function $\cg$ presented in Figure \ref{all_136}a
represents the integrated probability that a sight line through a halo
yields an absorber.  We note that this is governed by two physical
quantities: the mean covering fraction of cold gas per halo and the
fraction of halos that contain cold gas.  Interpretations of the
best-fit function must take these factors into account.

We find that $\cg$ increases rapidly with halo mass from being
vanishingly small at $M\sim 10^{10}$ \hmsol\ to unity by $M\sim
10^{11.5}$ \hmsol, at the midpoint of the cold-hot phase transition
(cf.\ Figure \ref{all_136}b). The strong constraints on the low-mass
cutoff in $\cg$ are driven primarily by the bias data. Figure
\ref{constraints} compares the bias data (points) and best-fit
transition model (solid curve) from Figure \ref{stats}c with an
additional model that demonstrate these constraints. The dotted curve
is the bias produced when low-mass halos contribute significantly to
the absorber statistics. This was calculated by adjusting the best-fit
model such that the minimum value of $\cg$ is 0.1, while reducing the
amplitude of $\cg$ at $M>10^{11}$ \hmsol\ by $\sim 30\%$ to correctly
match $\fr$. A contribution from 10\% of low-mass halos nearly washes
out the anti-correlation, but more importantly lowers the overall bias
of absorbers well below the measurements.

In the best fit model, the contribution at lower masses ($M\lesssim
10^{10.5}$) becomes negligible in order to match the clustering
bias. The numerical models of \cite{dekel_birnboim:06} predict that
the cold mode of accretion dominates in a range of halo masses rather
than a single transition. Shock heating, which is efficient for
massive halos, is also stable at $M\lesssim 10^{10}$\hmsol. Although
the virial temperatures at these mass scales are low, efficient shock
heating of halo gas in these halos could explain the dearth of
absorbers at low masses at $z\sim 0.6$.

However, the model indicates a small contribution to the total
absorbing cross section from $M\sim 10^{11}$ \hmsol\ halos. This can
be interpreted as either these halos have on average a few percent
covering fraction of warm/cold gas (producing the observed
\mg2\ features), or a few percent of galaxies in these halos have
extended Mg\,II gaseous halos.  An unbiased galaxy--Mg\,II pair
sample, established based on known galaxies along the lines of sight
to background QSOs discussed in \S\ 5.1, is necessary to break the
degeneracy between the two competing factors.

\cite{tripp_bowen:05} have recently measured the incidence of
\mg2\ absorption around galaxies to be $\sim 50\%$ out to 42
\hkpc\ (physical) for a sample of objects within the redshift range
$0.3-0.55$. This radius is roughly comparable to $\rg$ for $10^{11.5}$
\hmsol\ halos, in the regime where $\cg\approx 1$ in the best-fit
model. The gas radius in the model scales as $M^{1/3}$, and will
therefore change with the luminosity of the targeted galaxy. The range
in luminosities of the Tripp \& Bowen sample is 3 magnitudes, which
probes nearly an order of magnitude in halo mass (see, e.g., the HOD
analysis of SDSS and DEEP2 galaxy clustering data in
\citealt{zheng_etal:07}).  An incidence of less than unity is not
surprising given the large search radius and luminosity spread in the
\cite{tripp_bowen:05} study. Given the distribution of galaxy
luminosities in such a sample and a model for the halo occupation of
{\it galaxies} at the proper redshifts, a quantitative comparison can
be made and observational data on the incidence of absorption can be
used to further constrain the models.

At $M > 10^{12}$ \hmsol, where $\rshock = \rg$, the unity of $\cg$
indicate that all halos must contribute to the known Mg\,II statistics
through the cold gas that survived the shocks.  This explains the
overall strong bias of absorbers measured in B06, i.e. the elevated
mean bias is due to a significant contribution from high mass
halos. B06 have interpreted the anti-correlation between $\wr$ and
$M$ as due to predominant contribution to strong Mg\,II absorbers
($\wr>2$ \AA) from starburst outflows in low-mass systems. In our
model, however, this anti-correlation results from high mass halos
in the post-shock regime that have low cold gas fractions and can only
contribute to $\wr\lesssim 1$ \AA\ absorbers. 

Starbursts are not required to account for the mean mass of $\wr\sim
2$ \AA\ systems; absorption line with a width of 2 \AA\ implies at
maximum velocity difference of $\sim 200$ \kms\ between individual
components, which is less than twice the virial velocity of a
$10^{11.5}$ \hmsol\ NFW halo, 220 \kms. Due to the high covering
factors required to fit $\fr$, in the starburst scenario nearly all
halos of $10^{11.5-12}$ \hmsol\ would be experiencing an episode of
star formation strong enough to push outflows to one third the virial
radius.  We note, however, that outflows are expected to be
asymmetric, making it difficult to create high covering fractions
through starbursting alone.  Although rapidly star forming systems are
likely to contribute to absorber statistics to some degree, it is not
necessary to invoke a significant frequency of such systems to
reproduce the observations.

\subsection{Cold Gas in the Post-Shock Regime}

The constraints on $\fcold$ in the transition model are driven by a
combination of the frequency and bias data. If cold gas is too
abundant at high masses, then a $\bb$-$\wr$ anti-correlation cannot be
produced (as expected from the results of the classical model) and the
frequency distribution becomes tilted to higher $\wr$. The dashed
curve in Figure \ref{constraints} demonstates the effect of having no
cold gas in hot-mode halos by setting $\fcold=0$.  This weakens the
$\bhat$-$\wr$ anti-correlation has a strong effect on the overall bias
scale of the absorbers.

In our model $\fcold$ is independent of halos mass, thus a Milky-Way
size halo, just outside the transition regime, is predicted to have
only $\sim 6\%$ cold gas. Multiphase models of the gaseous halo by
\cite{maller_bullock:04} and \cite{chelouche_etal:07} predict high
cold gas fractions ($\gtrsim 0.2$) , but these predictions depend
sensitively on the adopted cloud size and are nearly an order of
magnitude more than observational inventories of the cold gas in the
Milky Way halo (\citealt{putman:06}).  The simulations of
\cite{keres_etal:05} and \cite{kravtsov:03} (as analyzed in
\citealt{birnboim_etal:07}) also imply $\fcold<0.1$ for all halos
$M>10^{12}$ \hmsol, but they do not have enough volume to address mass
dependence of $\fcold$.  With better bias data, it will be possible to
add an additional degree of freedom in the model and make $\fcold$
mass dependent.

Despite the poor constraints on the gas covering factor in the
post-shock regime ($M>10^{12.5}$ \hmsol), it is clear the bias data
are best fit by models with a high $\cg$ but a low cold gas
fraction. At the cluster scale, $M\gtrsim 10^{14}$ \hmsol, the virial
temperature of such halos is too high to support significant amounts
of low-ionization species. The situation at the group scale,
$10^{13-14}$ \hmsol, is less clear. The simulations of
\cite{keres_etal:05} and \cite{birnboim_etal:07} do not have
sufficient volume to probe this mass scale. A relevant question is
whether the two-phase medium of \cite{mo_jordi:96} and
\cite{maller_bullock:04} can exist at these halo masses. These authors
demonstrate that cold clouds can survive in massive halos, if these
clouds are massive themselves. \cite{dekel_birnboim:07} propose that
accreted cold clouds in massive halos can also be long-lived.
Observational searches for cold gas clouds in poor galaxy groups have
yielded null detections (\citealt{zabludoff:03}), although the
large-scale bias of Mg\,II absorbers requires that some fraction of
absorbers exist at this mass scale. Some cold gas in group and
cluster-sized halos should be associated with the satellite galaxies
contained within it, but the covering fraction of these satellites is
likely much less than unity.

Recent clustering measurements of $\lya$ absorbers by
\cite{ryan-weber:06} also yield a bias consistent with $\sim 10^{14}$
\hmsol\ halos. The cross-correlation function between LRGs and
absorbers itself can directly address this question.  If pairs exist at
separations less than $1$ \hmpc, then it demonstrates that massive
galaxies and cold gas exist in the same halo.  The cross-correlation
measurements of \cite{bouche_etal:04} show a signal at these small
separations, but the significance of the detection is low and should
be verified with a larger statistical sample.

\subsection{Toward Understanding the Origin of \mg2\ Absorbers}

We have demonstrated both that the halo-based approach can accurately
model the data and that the parameters of the model can be
well-constrained with those data.  More concrete physical inferences
about the distribution of cold gas in halos and its implications
require several improvements to the model and the data in forthcoming
studies.

%In particular, the present analysis is based on the approximation that
%the properties of dark matter halos are calculated at the effective
%redshift of the absorber sample. As shown in \S\ 5, the gas content of
%halos must be evolving in some way to counter-balance the redshift
%evolution in the mass function itself.  To explain the mild redshift
%evolution in the number density of Mg\,II absorbers, a possible
%scenario is that the mean covering fraction of cold gas $\cg$ remains
%fixed in individual halos as the halos grow in mass.  This would
%suggest a higher gas accretion rate at higher redshift for a fixed
%mass. This has been seen in hydrodynamic simulations (e.g,
%\citealt{keres_etal:05}) as well as collisionless simulations tracking
%dark halo growth (\citealt{wechsler_etal:02, vdb:02,
%  neistein_etal:06}). Therefore, the evolving $\cg$ may be adopted to
%constrain the cold gas accretion versus time and performing this halo
%occupation analysis in multiple redshift bins can quantify that
%evolution.

Our analysis also demonstrates that the statistics of \mg2\ absorbers
can be modeled by a representative sample of the dark halo
population. The high incidence and covering factors required to fit
the data imply that QSO absorbers are a ubiquitous phenomenon of the
overall galaxy population. Galaxy-absorber pairs are not skewed to a
special subset of the galaxy population, such as starburst systems, or
biased to halos in a specific environment.

Additional data, such as an unbiased sample of galaxy-absorber pairs,
the line-of-sight autocorrelation of absorbers, and the correlations
between $\wr$ and associated galaxy properties will allow better
constraints in the model. Quantifying the relationship between cold
gas and dark matter halos is a key step in painting a more complete
picture of galaxy formation.  By tracking the baryons that do not
shine, the information obtainable through this approach is
complementary to that provided by observations of the galaxies
themselves.

%%%%%%%%%%%%%%%%%%%%%%%%%%%%%%%%%%%%%%%%%%%%%%%%%%%%%%%%%%%%%%%%%%%%%%%%
%  Acknowledgments
%%%%%%%%%%%%%%%%%%%%%%%%%%%%%%%%%%%%%%%%%%%%%%%%%%%%%%%%%%%%%%%%%%%%%%%%
%\vspace{1cm}

\acknowledgements

It is a pleasure to thank A.\ Kravtsov and B.\ Robertson for important
discussions during the development of this project, and D.\ Weinberg
for helpful comments on an earlier version of this paper.  The authors
thank J.\ Prochaska for providing updated measurements on the Mg\,II
frequency distribution function measurements prior to publication.
J.L.T. and H.-W.C. were partially supported by NASA grant NNG06GC36G
and NSF grant AST-0607510.

%%%%%%%%%%%%%%%%%%%%%%%%%%%%%%%%%%%%%%%%%%%%%%%%%%%%%%%%%%%%%%%%%%%%%%%%
%  Bibliography
%%%%%%%%%%%%%%%%%%%%%%%%%%%%%%%%%%%%%%%%%%%%%%%%%%%%%%%%%%%%%%%%%%%%%%%%

\bibliography{../risa}

%%%%%%%%%%%%%%%%%%%%%%%%%%%%%%%%%%%%%%%%%%%%%%%%%%%%%%%%1%%%%%%%%%%%%%%%
% FIGURES FIGURES 
%%%%%%%%%%%%%%%%%%%%%%%%%%%%%%%%%%%%%%%%%%%%%%%%%%%%%%%%%%%%%%%%%%%%%%%

\end{document}